%% file: paper.tex
\documentclass[natbib]{svjour3}                     
\smartqed  
\usepackage{graphicx,bm,times}
\usepackage{aps-bibstyle}  
\graphicspath{{./fig/}{./png/}}

\graphicspath{{./fig/}{./png/}}

\input{macros}

\begin{document}

\title{Astrophysical hydromagnetic turbulence}
\titlerunning{Astrophysical hydromagnetic turbulence}

\author{A. Brandenburg \and A. Lazarian}
\authorrunning{A. Brandenburg and A. Lazarian}

\institute{A. Brandenburg \at
Nordita, KTH Royal Institute of Technology and Stockholm University,
Roslagstullsbacken 23, 10691 Stockholm, Sweden; and
Department of Astronomy, Stockholm University, SE 10691 Stockholm, Sweden,
\email{brandenb@nordita.org}
\and A. Lazarian \at
Department of Astronomy, University of Wisconsin-Madison \\
475 N. Charter St., Madison, WI 53706, USA
\email{lazarian@astro.wisc.edu}
}

\date{\today,~ $ $Revision: 1.53 $ $}

\maketitle

\begin{abstract}
Recent progress in astrophysical hydromagnetic turbulence is being reviewed.
The physical ideas behind the now widely accepted Goldreich--Sridhar model and
its extension to compressible magnetohydrodynamic turbulence are introduced.
Implications for cosmic ray diffusion and acceleration is being discussed.
Dynamo-generated magnetic fields with and without helicity are contrasted
against each other.
Certain turbulent transport processes are being modified and often suppressed
by anisotropy and inhomogeneities of the turbulence, while others are
being produced by such properties, which can lead to new large-scale
instabilities of the turbulent medium.
Applications of various such processes to astrophysical systems are
being considered.
\keywords{magnetic fields \and turbulence \and Sun: magnetic fields
\and ISM: magnetic fields}
\PACS{44.25.+f\and 47.27.Eq\and 47.27.Gs \and 47.27.Qb}
\end{abstract}

\section{Introduction}
\label{intro}

Hydromagnetic or magnetohydrodynamic (MHD) turbulence plays an important
role in many astrophysical settings.
In a recent review by \cite{BN11}, properties of turbulence were discussed
for the solar wind, stellar convection zones, the interstellar medium,
accretion discs, galaxy clusters, and the early Universe.
In an earlier review by \cite{BS05}, a detailed account of dynamo theory
with emphasis on helical dynamos was given.
In that review, and also in \cite{BSS12}, the small-scale dynamo was
discussed in detail.
Applications to galactic dynamos were discussed by \cite{Beck}.
Aspects of magnetic reconnection and particle acceleration in turbulent
flows have recently been reviewed by \cite{Laz12}.
In the present review we begin with turbulence in the interstellar medium,
discuss how turbulence is affected by magnetic fields and compressibility,
address then applications to cosmic ray scattering and turn then to
dynamo-generated magnetic fields as well as to anisotropic and
inhomogeneous flows that are affected by stratification and rotation.

\section{Turbulence in the interstellar medium}
\label{BigPowerLaw}

\begin{figure}
\centering
\includegraphics[height=.65\textheight]{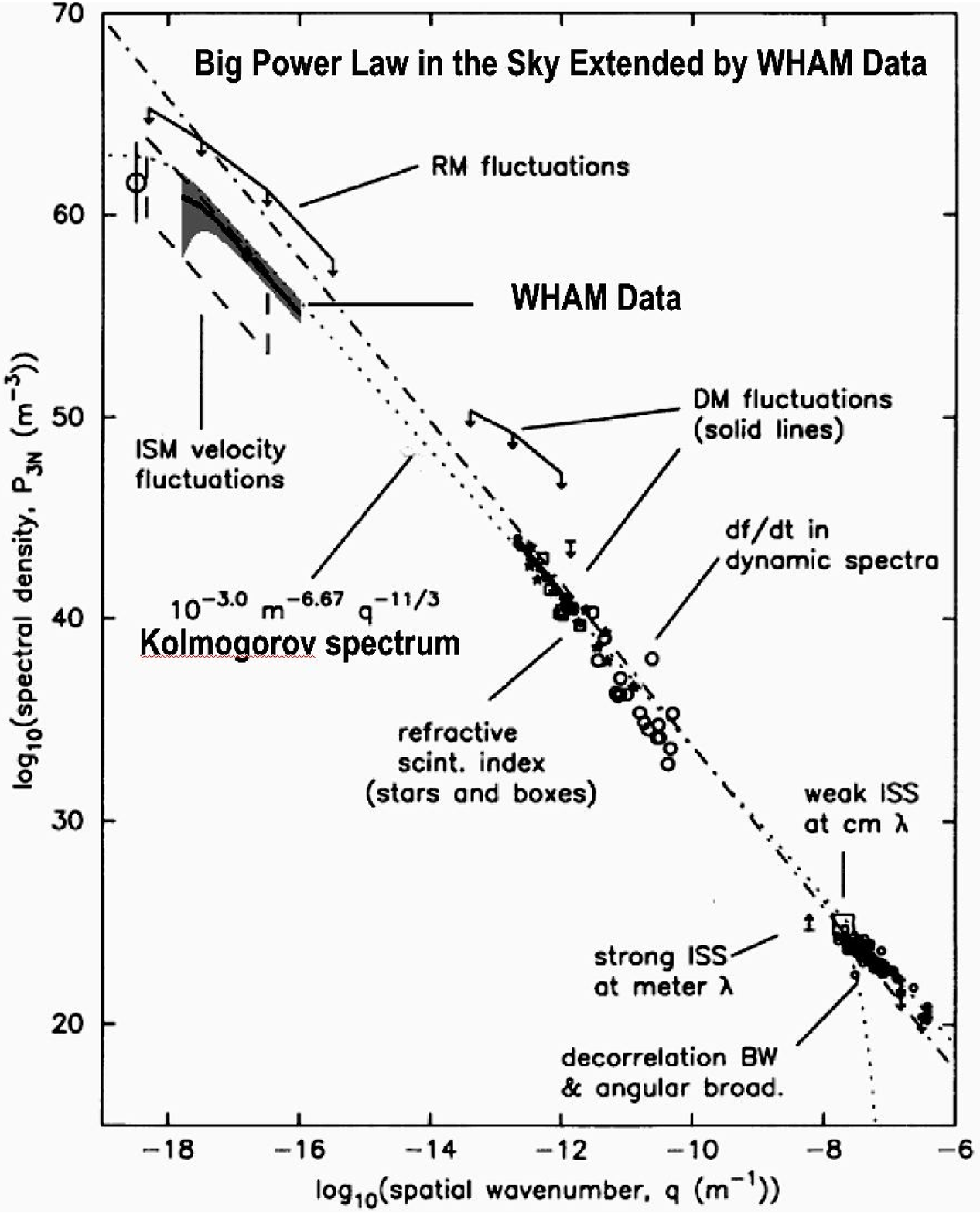}
\caption{Turbulence in the interstellar gas obtained from electron
density fluctuations.
The ``Big Power Law in the Sky'' of \cite{Arm95} is here
extended using data from the Wisconsin H$\alpha$ Mapper (WHAM).
The slope corresponds to that of Kolmogorov turbulence.
Adapted from \cite{CL10}.
}\label{CL}
\end{figure}

The ISM is turbulent on scales ranging from AUs to kpc
\citep{Arm95,ES04}, with an embedded magnetic field
that influences almost all of its properties.
MHD turbulence is accepted to be of key importance
for fundamental astrophysical processes, e.g.\ star formation,
propagation and acceleration of cosmic rays.
It is therefore not surprising that attempts to obtain spectra
of interstellar turbulence have been numerous since the 1950s \citep{Mun58}.
However, various directions of research achieved varying degrees of success.
For instance, studies of turbulence statistics of ionized media
accompanied by theoretical advancements in understanding scattering
and scintillation of radio waves in ionized media \citep{GN85}
were rather successful \citep[cf.][]{SG90}.
This work provided information about the statistics of the electron density
on scales $10^{8}$--$10^{15}$\,cm \citep{Arm95}.
These measurements have been recently combined with data from the
Wisconsin H$\alpha$ Mapper, which also measures electron density
fluctuation, but on larger scales.
The resulting extended spectrum presented in \cite{CL10}
shows that the Kolmogorov $-5/3$ spectrum of electron density fluctuations
extends to several more decades to larger scales; see \Fig{CL}.

In spite of their success, these sort of measurements provide only
density statistics, which is a rather indirect measure of turbulence. 
Velocity statistics is a much more direct turbulence measure. 
Although it is clear that Doppler broadened lines
are affected by turbulence, recovering the velocity statistics is
extremely challenging without adequate theoretical insight.
Indeed, both the $z$ component of velocity and density
contribute to fluctuations of the energy density
$\rho_{\rm s} ({\bm X}, V_z)$ in Position-Position-Velocity (PPV) space.

Traditionally, information on turbulence spectra is obtained using
the measure of Doppler shifts termed Velocity Centroids,
$\sim \int V_z \rho_{\rm s} dV_z$, where the integration is taking place
over the range of the velocities relevant to the object under study.
In this situation it is easy to see that the Velocity Centroids are
also proportional to $\int V_z \rho ds$, where $\rho$ is the actual
three-dimensional density and the integration is performed along the
line of sight \citep{LE03}.

Usually the Velocity Centroids are normalized by the intensity integrated
over the line of sight \citep{Ste90}, and the work of \cite{LE05}
showed that this normalization does not change the statistical
properties of the measure.
However, the numerical and analytical analysis in \cite{LE05}
and \cite{Esq07} showed that the Velocity Centroids fail for
studying supersonic turbulence.
This provides bad news for the studies of velocity statistics in
molecular clouds and the diffuse cold ISM \citep{DK85,MSB99,Miv03}
The studies for HII regions \citep{OdC87} are less strongly affected,
as in most cases the turbulence there is subsonic.

There have been attempts to analyze PPV data cubes in other ways.
For instance, \cite{CD83}, \cite{Gre93}, and \cite{Sta99}
analyzed power spectra of velocity channels of HI data.
The spatial spectrum of fluctuations of these velocity slices of PPV
revealed power-law dependences, but the physical meaning of these
dependences remained unclear.
(Indeed, some of the authors erroneously identified the spectral index
of intensity perturbations in slices of PPV data with the spectral index
of the underlying turbulence spectrum.
The nature of the variations of the spectral index in different studies
was unclear.)

The analytical study of the statistical properties of the PPV energy
density $\rho_{\rm s}$ has been initiated by \cite{LP00}.
There the observed statistics of $\rho_{\rm s}$ was related to the underlying
3D spectra of velocity and density in the astrophysical turbulent
volume.
Initially, the volume was considered transparent, but later the
treatment was generalized to volumes with self-absorption and to
studies of turbulence using absorption lines \citep{LP04,LP06,LP08}.
The resulting theory of mapping of fluctuations in
Position-Position-Position space with turbulent velocity into PPV space
was successfully tested in a number of studies
\citep{PJKN06,PJKN09,CL09,Burkhart13}.
This theory lays the foundation for two separate techniques, Velocity
Channel Analysis (VCA) and Velocity Correlation Spectrum (VCS) which
were applied by a number of groups to different data sets including
HI, C O$^{13}$, $^{18}$C O; see more in \cite{Laz09}.
The results can be briefly summarized as follows: the tested supersonic
media exhibit a velocity spectrum that is steeper than
the spectrum of Kolmogorov turbulence and a density spectrum
that is shallower.
This result is, in fact, expected for supersonic MHD turbulence
\citep{BLC05,KLB07}.

We emphasize that VCA and VCS are two related techniques based on solid
analytical foundations.
The theory of the VCA in \cite{LP00,LP04}
and VCS in \cite{LP06,LP08} describe the non-linear
mapping provided by velocity fluctuations from the turbulent volume to the
Position-Position-Velocity (PPV) space. Therefore the technique provides
the true spectrum of velocity and density fluctuations, irrespectively
of the sources and sinks of turbulence. The energy injection associated
with localized injection of turbulence, e.g. with the outflows should
be detected as the changes in the spectral slope corresponding to the
scales of energy injection.

\section{The picture of Alfv\'enic Turbulence}

The picture of MHD turbulence has been developing over decades
and pioneering works by \cite{Iro63} and \cite{Kra65} are
definitely to be mentioned. The Iroshnikov-Kraichnan model was the
extension of Kolmogorov's {\it isotropic} turbulence model and it is
the assumption of anisotropy that was a deficiency of this model.
The notion of anisotropic turbulence was established later in important
works, notably, by \cite{SMM83} for incompressible turbulence
and \cite{Hig84} for the compressible turbulence.
These papers provided the ground for the further advance.

Quantitative insight into MHD turbulence has been obtained in
the seminal paper by \cite{GS95}, hereafter referred to as GS95.
This paper quantified the properties of the anisotropic cascade and
provided foundations for further theoretical development in the field.
We may mention parenthetically that the original paper could not provide
the perfect picture of MHD turbulence theory and a number of key
aspects were clarified and corrected in subsequent studies.
For instance, the original claim in GS95 and \cite{GS97} about the
role of 3-wave interactions were later corrected, and for weak MHD
turbulence the point of view expressed in \cite{NB96} was adopted.
Similarly, the notion of a {\it local} system of reference
that is essential for understanding critical balance, which is
a corner stone of our modern understanding of GS95 theory, was missing
in the original paper.
In fact, the closure relations that are used in GS95 to justify the model
are written in the system of reference related to the mean field
and therefore cannot be used as a proof.
The importance of a local system of reference was understood only in
subsequent theoretical and numerical studies by \cite{LV99}, henceforth LV99,
\cite{CV00}, as well as \cite{MG01}.

\subsection{Incompressible MHD turbulence}

While having a long history of ideas, the theory of MHD turbulence has
become testable recently with the advent of numerical simulations
\citep{Bis03}, which confirmed \citep[see][and references therein]{CL05}
the prediction of magnetized Alfv\'enic eddies being elongated
in the direction of the magnetic field \citep{SMM83,Hig84}
and provided results consistent with quantitative
relations for the degree of eddy elongation obtained by GS95.

MHD turbulence theory is in many respects similar to the famous
\cite{Kol41} theory of turbulence. In the latter theory, energy
is injected at large scales, creating large eddies which
do not dissipate energy through
viscosity\footnote{The Reynolds number
$\Rey\equiv L_{\rm f}V/\nu=(V/L_{\rm f})/(\nu/L^2_{\rm f})$
characterizes the ratio of the eddy turnover rate
$\tau^{-1}_{\rm eddy}=V/L_{\rm f}$ and the viscous dissipation rate
$\tau_{\rm dis}^{-1}=\eta/L^2_{\rm f}$. Therefore large values of
$\Rey$ correspond to negligible viscous dissipation of large eddies over the
cascading time $\tau_{\rm casc}$ which is equal to $\tau_{\rm eddy}$ in
Kolmogorov turbulence.} but transfer energy to smaller eddies.
The process continues
until the cascade reaches the eddies that are small enough to dissipate
energy over an eddy turnover time. In the absence of compressibility the
hydrodynamic cascade of energy is $\sim v^2_l/\tau_{{\rm casc}, l} =\const$,
where $v_l$ is the velocity at the scale $l$ and the cascading time for
the eddies of size $l$ is $\tau_{{\rm casc}, l}\approx l/v_l$. From this the
well known relation $v_l\sim l^{1/3}$ follows.

In MHD turbulence, in the presence of dynamically important magnetic
fields, eddies become anisotropic. At the same time, one can imagine
eddies mixing magnetic field lines perpendicular to the direction
of the magnetic field. For these eddies, the original Kolmogorov
treatment is applicable resulting in perpendicular motions scaling as
$v_l\sim \l_{\bot}^{1/3}$, where $l_{\bot}$ denotes eddy scales measured
perpendicular to the magnetic field. These mixing motions induce Alfv\'enic
perturbations that determine the parallel size of the magnetized eddy.
A cornerstone of the GS95 theory is {\it critical balance}, i.e. the
equality of the eddy turnover time $l_{\bot}/v_l$ and the period of
the corresponding Alfv\'en waves $\sim l_{\|}/V_{\rm A}$, where $l_{\|}$ is
the parallel eddy scale and $V_{\rm A}$ is the Alfv\'en velocity. Making use
of the earlier expression for $v_l$, one can easily obtain $l_{\|}\sim
l_{\bot}^{2/3}$, which reflects the tendency of eddies to become more
and more elongated as energy cascades to smaller scales.

It is important to stress that the scales $l_{\bot}$ and $l_{\|}$ are
measured with respect to a system of reference related to the direction
of the local magnetic field ``seen'' by the eddy. This notion was not
present in the original formulation of the GS95 theory and was added
to it later by \cite{LV99}, henceforth LV99, and \cite{CV00}.
The local system of reference was also used in numerical studies in
\cite{CV00}, \cite{MG01}, and \cite{CLV02} that tested GS95 theory.
In terms of mixing motions, it is rather obvious that the free
Kolmogorov-type mixing is possible only with respect to the local magnetic
field of the eddy rather than the mean magnetic field of the flow.

While the arguments above are far from being rigorous, they correctly
reproduce the basic scalings of magnetized turbulence when the velocity
is equal to $V_{\rm A}$ at the injection scale $L$. The most serious argument
against this picture is the ability of eddies to perform mixing motions
perpendicular to the magnetic field. This problem was addressed in LV99,
where the self-consistency of the GS95 theory was related
to fast reconnection of the magnetic field in turbulent fluids.
A more rigorous discussion of a self-consistent treatment of MHD
turbulence and magnetic reconnection is presented in \cite{ELV11}.

\begin{table*}[t]
\caption{Regimes and ranges of MHD turbulence}
\begin{tabular}{lllll}
\hline
Type of MHD turbulence & Injection velocity & Range of scales &
Motion type & Ways of study\\
\hline
Weak & $V_{\rm L}<V_{\rm A}$ & $[L, l_{\rm trans}]$ & wave-like & analytical\\
Strong subAlfv\'enic & $V_{\rm L}<V_{\rm A}$ & $[l_{\rm trans}, l_{\min}]$ & eddy-like & numerical \\
Strong superAlfv\'enic & $V_{\rm L}> V_{\rm A}$ & $[l_{\rm A}, l_{\min}]$ & eddy-like & numerical \\
\hline
\\
\multicolumn{5}{l}{\footnotesize{$L$ and $l_{\min}$ are injection and dissipation scales}}\\
\multicolumn{5}{l}{\footnotesize{$l_{\rm trans}$ and $l_{\rm A}$ are given
by \Eqs{trans}{alf}, respectively.}}\\
\label{Regimes}
\end{tabular}
\end{table*}

The GS95 theory is formulated assuming isotropic injection of energy at
scale $L$ and the injection velocity equal to the Alfv\'en velocity in
the fluid $V_{\rm A}$, i.e.\ the Alfv\'en Mach number $M_{\rm A}\equiv (V_{\rm L}/V_{\rm A})=1$,
where $V_{\rm L}$ is the injection velocity.
Thus, it provides the description of transAlfv\'enic turbulence.
This model was later extended for both subAlfv\'enic, i.e.\ $M_{\rm A}<1$,
and superAlfv\'enic, i.e. $M_{\rm A}>1$, cases
(see LV99 and Lazarian 2006, respectively; see also \Tab{Regimes}).
Indeed, if $M_{\rm A}>1$, then, instead of the driving scale $L$ one can use
the scale
\begin{equation}
l_{\rm A}=LM_{\rm A}^{-3},
\label{alf}
\end{equation}
which is the scale at which the turbulent velocity equals $V_{\rm A}$.
For $M_{\rm A}\gg 1$, magnetic fields are not dynamically important at the
largest scales and the turbulence at those scales follows the isotropic
Kolmogorov cascade $v_l\sim l^{1/3}$ over the range of scales $[L, l_{\rm A}]$.
At the same time, if $M_{\rm A}<1$, the turbulence obeys GS95 scaling (also
called ``strong'' MHD turbulence) not from the scale $L$, but from a
smaller scale
\begin{equation}
l_{\rm trans}=LM_{\rm A}^2,
\label{trans}
\end{equation}
while in the range $[L, l_{\rm trans}]$ the turbulence is ``weak''. 

The properties of weak and strong turbulence are rather different.
Weak turbulence is wave-like turbulence with wave packets undergoing
many collisions before transferring energy to small scales.
Unlike strong turbulence, weak turbulence allows an exact analytical
treatment \citep{Gal00}.
By contrast, in strong turbulence intensive interactions between
wave packets prevent the use of a perturbative approach.
Numerical experiments have supported the GS95 ideas both for
incompressible MHD turbulence \citep{CV00,MG01,CLV02,BL10,Ber11}
and for the Alfv\'enic component of compressible MHD turbulence
\citep{CL02,CL03,KL10}.
[The compressible MHD turbulence simulations of \cite{BLC05} and
\cite{KLB07} demonstrated that the density spectrum becomes more shallow
and isotropic as the Mach number increases.]

While there are ongoing debates whether the original GS95 theory must
be modified to better describe MHD turbulence, we believe that we do not
have compelling evidence that GS95 is not adequate.
The most popular one is the modification of the GS95 model by
\cite{Bol05,Bol06}, who, motivated by the spectral index of $-3/2$ observed
in simulations of \cite{MG01}, proposed that the difference of the
GS95 predictions and the numerical experiments arises from the dynamical
alignment of velocity and magnetic fields.
However, \cite{BL09,BL10} showed that present day
numerical simulations may not have enough resolution to reveal the actual
inertial range of MHD turbulence and the existing numerical simulations
may be dominated by the bottleneck effect that distorts the actual slope
of turbulence. Incidentally, the bottleneck effect already played a
trick with the researchers when supersonic simulations suggested a $-5/3$
spectrum of supersonic turbulence \citep{BNP02} which later
was proven to be a bottleneck effect of shock wave turbulence with the
expected $-2$ spectrum \citep{Kritsuk}.
Such a spectrum has been confirmed with several different codes
\citep{Kritsuk11}. 
In addition, the $-5/3$ spectral index agrees well with the resolution
studies by \cite{Ber11,Ber12}.
Thus, within the present review we will refer to GS95 when we shall talk
about strong MHD turbulence.

The issue of the spectral slope is of both theoretical and practical
importance. Although the differences between spectral slopes of 5/3
and 3/2 or even 2 do not look large, they correspond to very different
physical pictures. The spectrum of 3/2 corresponds to interactions
decreasing with the scale of turbulent motions, 5/3 corresponds to
a strongly Kolmogorov-type picture of eddies, while 2 corresponds to
a spectrum of shocks. The anisotropies predicted in these different
pictures of turbulence are also different.
They are in fact extremely important for cosmic ray propagation;
see \cite{YL04} and references therein.
We also note that even a small difference in the slope
can result in substantial differences in the energy at small scales due
to the enormous extent of the astrophysical turbulent cascade.
Finally, as GS95 has now the status of the accepted model of turbulence,
it is essential to test all the predictions of this theory, including
the predicted 5/3 spectral slope.

Usually, one considers balanced turbulence, i.e.\ the situation
when the flows of energy in opposite directions are equal.
In a more general case the turbulence is imbalanced, i.e.\ the flow of
energy from one side dominates the flow from the opposite direction.
The existing models of imbalanced turbulence are hotly debated at the
moment and their predictions are being tested \citep{LGS07,BL08,PB09}.
Here we will just mention that in the case of astrophysical turbulence,
compressibility may decrease the degree of imbalance, making the simple
GS95 model applicable in spite of the presence of sources and sinks
of energy.

\subsection{Compressible MHD turbulence}
\label{CompressibleMHDturbulence}

\begin{figure*} 
\begin{center}
\includegraphics[width=0.8\textwidth]{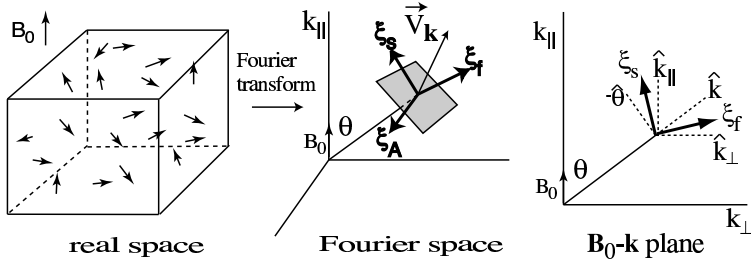}
\end{center}
 \caption{Graphical representation of the mode separation method.
We separate the Alfv\'{e}n, slow and fast modes by the projection
of the velocity Fourier component ${\bf v}_k$ on the bases
${\bf \hat\xi}_{\rm A}$, ${\bf \hat\xi}_{\rm s}$ and
${\bf \hat\xi}_{\rm f}$, respectively.
Adapted from CL03.\label{Cho_Laz}}
\end{figure*}

The statistical decomposition of 3D MHD turbulence into fundamental modes,
i.e.\ Alfv\'en, slow and fast, was performed in Fourier space by
\cite{CL02,CL03}, henceforth CL02 and CL03, respectively, and later
using wavelets by \cite{KL10}.
The idea of the decomposition is presented in Figure~\ref{Cho_Laz}.
The procedure was tested with the decomposition in real space in special
cases when such a decomposition was possible, for instance, in the case
of slow modes in a low plasma-$\beta$ medium.

\begin{figure*}
\includegraphics[width=0.99\textwidth]{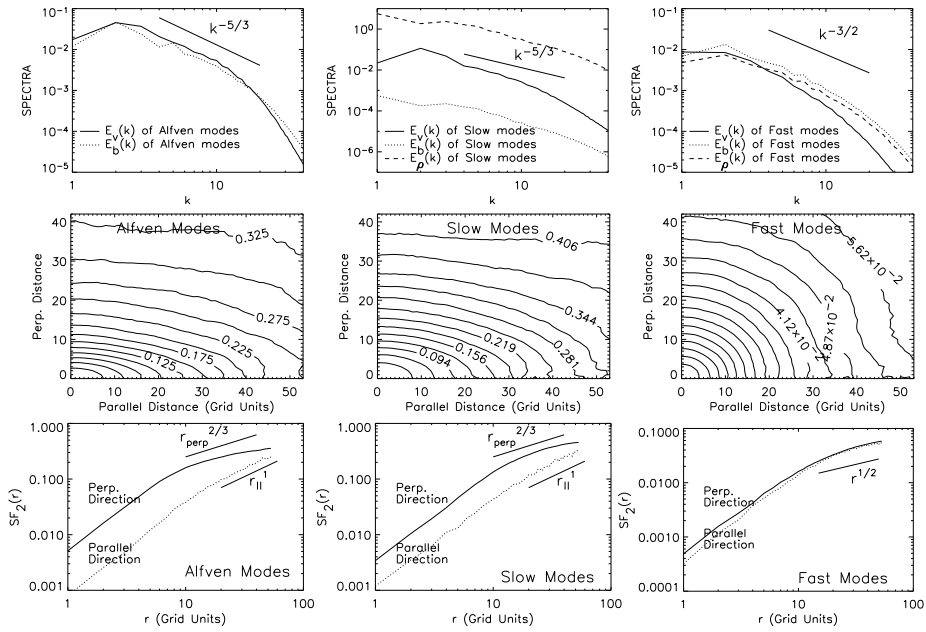}
\caption{
Highly supersonic low $\beta$ ($\beta\sim0.02$ and $M_{\rm s}\sim$7). 
$V_{\rm A}\equiv B_0/\sqrt{4\pi\rho}=1$. $a$ (sound speed) $=0.1$. 
$\delta V \sim 0.7$.
Alfv\'en modes follow the GS95 scalings. Slow modes follow
the GS95 anisotropy. 
Fast modes are isotropic.
}\label{f2ch}
\end{figure*}

The most important result of this decomposition was establishing
the relevance of Alfv\'enic turbulence scaling to a compressible medium.
As we see in \Fig{f2ch}, the anisotropy of the Alfv\'enic
component corresponds to the GS95 predictions.
In general, the study of transAlfv\'enic turbulence with different Mach
numbers in CL02 and CL03 revealed that GS95 scaling is valid for
{\it Alfv\'en modes}:
$$
   \mbox{ Alfv\'{e}n:~}  E^{\rm A}(k)  \propto k^{-5/3}, 
                        ~~~k_{\|} \propto k_{\perp}^{2/3}. 
$$
{\it Slow modes} also follow the GS95 model for both
high $\beta$ and mildly supersonic low $\beta$ cases:
$$
   \mbox{ Slow:~~~}   E^{\rm s}(k)  \propto k^{-5/3}, 
                        ~~~k_{\|} \propto k_{\perp}^{2/3}.  
$$
For the highly supersonic low $\beta$ case, the kinetic energy spectrum of 
slow modes tends to be steeper, which may be related
to the formation of shocks.\\
{\it Fast mode} spectra are compatible with
acoustic turbulence scaling relations:
$$
   \mbox{ Fast:~~~}   E^{\rm f}(k)  \propto k^{-3/2}, 
                        ~\mbox{isotropic spectrum}.   
$$

The coupling between fast and Alfv\'en modes was shown to be weak and
therefore the cascades of fast and Alfv\'en modes weakly affect each
other (CL02).
At the same time, Alfv\'en modes cascade to slow modes, which are
otherwise passive in the cascade.
This corresponds to the theoretical expectations discussed in GS95,
\cite{LG03}, and CL03.

In terms of energy transfer from Alfv\'enic to compressible modes CL02
suggested the theory-motivated expression
\begin{equation}
  \frac{\delta E_{\rm comp}}{\delta E_{\rm Alf}}\approx \frac{\delta V_{\rm A} V_{\rm A}}{V_{\rm A}^2+c_{\rm s}^2},
\label{eq_high2}
\end{equation}
where $\delta E_{\rm comp}$ and $\delta E_{\rm Alf}$ are the energies
of compressible  and Alfv\'en modes, respectively.
\Eq{eq_high2} suggests that the drain of energy from
Alfv\'enic cascade is marginal
when the amplitudes of perturbations
are weak, i.e.\ $(\delta V)_{\rm A} \ll  V_{\rm A}$.
Results of numerical calculations in CL02 are consistent with the
expression above.
The marginal transfer of energy between Alfv\'enic and compressible
motions justifies considering the Alfv\'enic and fast cascades separately.

Higher resolution simulations in \cite{KL10} used a different
wavelet-based decomposition technique.
The results agree well with those in CL03.
The advantage of the wavelet decomposition is the ability to decrease
the error for the case of strongly perturbed fields.

\section{Implications of MHD turbulence for diffusion processes}

\subsection{Diffusion of heat by MHD turbulence}

Transport processes are known to be affected by turbulence.
A big issue related to MHD turbulence is the nature of turbulent
eddies.
If magnetic field lines are perfectly frozen into fluid, then one cannot
talk about mixing motions at the scale of dynamically important magnetic
fields.
On the contrary, if magnetic reconnection is fast enough to resolve the
knots of intersecting magnetic fluxes that naturally arise in GS95
turbulence, mixing motions perpendicular to the local magnetic field
should be similar to those in hydrodynamical fluids.
This problem was addressed in LV99, where it was shown that
magnetic reconnection induced by turbulence makes the
GS95 picture of a perpendicular cascade self-consistent.
A more recent study by \cite{ELV11} revealed the
deep connection between turbulence and magnetic reconnection.
This provides a theoretical justification for discussing hydrodynamic-type
turbulent advection of heat in the presence of dynamically important
magnetic fields.\footnote{The arguments in \cite{ELV11} should be
distinguished from the arguments based on attempted renormalization of
the effective magnetic Reynolds numbers in \cite{BF08}.
\cite{ELV11} do not introduce artificial ``turbulent diffusivities''
but appeal to the established and tested concept of Richardson diffusion.}

In addition, in hot plasmas, the motion of electrons along wandering
magnetic fields is important.
The statistics of magnetic field wandering was described in LV99 for
different regimes of turbulence and provides the necessary foundations
for a quantitative description of the heat transfer process.
This is the process that we start our discussion with.

Let us initially disregard the dynamics of fluid motions on diffusion,
i.e.\ we consider diffusion induced by particles moving along wandering
turbulent magnetic field lines, whose motions we disregard for the sake of
simplicity.
Magnetized  turbulence with a dynamically important magnetic field is
anisotropic with eddies elongated along the direction of local magnetic
field (henceforth denoted by $\|$), i.e.\ $l_{\bot}<l_{\|}$, where
$\bot$ denotes the direction perpendicular to the local magnetic
field.
Consider isotropic injection of energy at the outer scale $L$ and
dissipation at the scale  $l_{\bot,\min}$.
This scale corresponds to the minimal dimension of the turbulent eddies.

Initially, the problem of heat transport by electrons moving in turbulent
magnetic fields was considered by \cite{NM01} for transAlfv\'enic turbulence.
The treatment for both  subAlfv\'enic and superAlfv\'enic turbulence
was presented in \cite{Laz06}; henceforth L06.

It is easy to notice that the separations of magnetic field lines at
scales below the damping scale of turbulence, i.e.\ for
$r_0<l_{\bot,\min}$, are mostly influenced by the motions at the smallest scale.
This scale $l_{\bot,\min}$ results in Lyapunov-type growth
$\sim r_0 \exp(l/l_{\|,\min})$.
This growth is similar to that obtained in earlier models with a single
scale of turbulent motions; see \cite{RR78}, henceforth RR78, and \cite{CC98}.
Indeed, as the largest shear that causes field line divergence is due
to the marginally damped motions at the scale around $l_{\bot,\min}$
the effect of larger eddies can be neglected and we are dealing with
the case of single-scale ``turbulence'' described by RR78.

The electron Larmor radius presents the minimal perpendicular scale of
localization, while the other relevant scale is the Ohmic diffusion scale
corresponding to the scale of damped motions.
Thus, conservatively it is natural to associate $r_0$ with the size of the
cloud of electrons of the electron Larmor radius $r_{\rm Lar, particle}$.
Applying the original RR78 theory, they found that the electrons should
travel over a distance
\begin{equation}
L_{\rm RR}\sim l_{\|,\min} \ln (l_{\bot,\min}/r_{{\rm Lar}, e})
\label{RR}
\end{equation}
to get separated by $l_{\bot,\min}$.

Within the single-scale ``turbulence model'', which formally corresponds
to $L{\rm ss}=l_{\|,\min}=l_{\bot,\min}$, the distance $L_{\rm RR}$ is called
Rechester--Rosenbluth distance.
For the intracluster medium parameters, for which the problem was discussed
originally, the logarithmic factor in \Eq{RR} is of the order of $30$.
This causes a $30$-fold decrease of the thermal conductivity for
the single-scale models\footnote{For the single-scale model, $L_{\rm RR}\sim
30L$ and the diffusion over distance $\Delta$ takes $L_{\rm RR}/L{\rm ss}$
steps, i.e.\ $\Delta^2\sim L_{\rm RR} L$, which decreases the corresponding
diffusion coefficient $\kappa_{e,{\rm single}}\sim \Delta^2/\delta t$ by a
factor 30.}.

The single-scale turbulence model is just a toy model to study the effects
of turbulent motions.
However, one can use this model to describe what is happening below the
scale of the smallest eddies.
Indeed, shear and, correspondingly, magnetic field line divergence are
maximal for the marginally damped eddies at the dissipation scale.
Thus, for scales less than the damping scale the action of the critically
damped eddies is dominant and the results of \Eq{RR} are applicable.
The additional traveling distance of $L_{\rm RR}$ is of marginal importance
for diffusion of heat over distances $\gg L_{\rm RR}$.

For the diffusion in superAlfv\'enic turbulence the Alfv\'enic scale $l_{\rm A}$
given by \Eq{alf} is important.
It acts as the characteristic scale of magnetic fluctuations.
Assuming that the mean free path of electrons is less than $l_{\rm A}$,
L06 obtained:
\begin{equation}
\kappa_{e}\equiv \Delta^2/\delta t\approx (1/3) l_{\rm A} v_{e},~~~ l_{\rm A}<\lambda,
\label{el}
\end{equation}
where $v_e$ is the electron velocity.
In the opposite limit of effective scattering $\lambda< l_{\rm A}$, we have
$\kappa \sim \lambda v_e$ with the coefficient of proportionality equal
to $1/5$ according to \cite{NM01}.

For subAlfv\'enic turbulence, the turbulence gets into
the regime of strong GS95 type turbulence, which is described by \Eq{trans}.
The diffusivity becomes anisotropic with the diffusion coefficient parallel
to the mean field, $\kappa_{\|, {\rm particle}}\approx 1/3 \kappa_{\rm unmagn}$
being larger than the coefficient for diffusion perpendicular to the magnetic
field (L06):
\begin{equation}
\kappa_{\bot, e}=\kappa_{\|, e}M_{\rm A}^4,~~~ M_{\rm A}<1,
\label{9}
\end{equation}
As discussed above, turbulent motions themselves can induce
advective transport of heat.
Appealing to the LV99 model of reconnection, one can conclude that turbulence
with $M_{\rm A}\sim 1$ should be similar to hydrodynamic turbulence, i.e.\
\begin{equation}
\kappa_{\rm dynamic}\approx C_{\rm dyn} L V_{\rm L},~~~ M_{\rm A}>1,
\label{dyn}
\end{equation}
where $C_{\rm dyn}\sim 0(1)$ is a constant, which for hydro turbulence is
around $1/3$ \citep{Les90}.
If we deal with heat transport, for fully ionized non-degenerate plasmas
we assume $C_{\rm dyn}\approx 2/3$ to account for the advective heat transport
by both protons and electrons.

The advection of heat in the regime of subAlfv\'enic turbulence is
reduced compared to the superAlfv\'enic case and given by expression
(L06):
\begin{equation}
\kappa_{\rm dynamic}\approx (\beta/3) LV_{\rm L} M_{\rm A}^3, ~~~ M_{\rm A}<1,
\label{dyn_weak}
\end{equation}
where $\beta\approx 4$.

\begin{figure}\centering
\includegraphics[width=.8\textwidth]{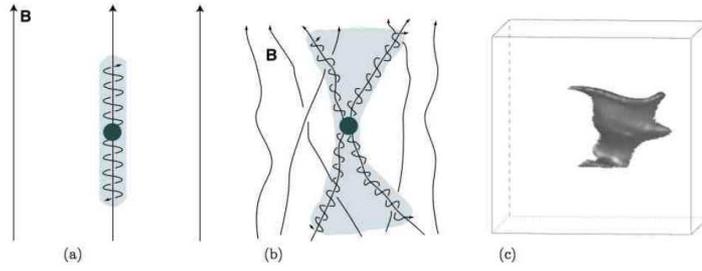}
\caption{\cite{EV06}.
(a) Textbook picture of electrons moving along magnetic field lines in thermal conduction process. (b) Actual motions of electrons in realistic turbulent plasmas where reconnection and spontaneous stochasticity of magnetic field are present. (c) Numerical simulations of heat advection in magnetized turbulence.
From \cite{CL04}. 
}\label{space}\end{figure}

Figure~\ref{space} illustrates the existing ideas on processes of
heat conduction in astrophysical plasmas.
They range from heat insulation by unrealistically laminar magnetic
field (a), to heat diffusion in turbulent magnetic field (b) and to heat
advection by turbulent flows (c).
field, to heat diffusion in turbulent magnetic field and to heat
advection by turbulent flows.
The relative efficiencies of the two latter processes depend on
parameters of the turbulent plasma.
The observational data for two clusters are also shown and it is clear
that for the clusters of galaxies discussed, the turbulent advection of heat
is the dominant process.
The dominance of turbulent motions gets even more prominent if one takes
into account that instabilities in the collisionless plasma of galaxies
are likely to dramatically decrease the mean free path of electrons.

In thermal plasma, electrons are mostly responsible for thermal conductivity.
The schematics of the parameter space for
$\kappa_{\rm particle}<\kappa_{\rm dynamic}$ is shown in \Fig{space}, where
the Mach number $M_{\rm s}$ and the Alfv\'en Mach number $M_{\rm A}$ are the variables.
For $M_{\rm A}<1$, the ratio of diffusivities arising from fluid and particle
motions is
$\kappa_{\rm dynamic}/\kappa_{\rm particle}\sim \beta\alpha M_S M_{\rm A}(L/\lambda)$;
see \Eqs{9}{dyn_weak}.
The square root of the ratio of the electron to proton mass
$\alpha=(m_e/m_p)^{1/2}$, which provides the separation line between
the two regions in Fig.~2, is given by $\beta\alpha M_{\rm s}\sim(\lambda/L) M_{\rm A}$.
For $1<M_{\rm A}<(L/\lambda)^{1/3}$ the mean free path is less than $l_{\rm A}$ which
results in $\kappa_{\rm particle}$ being some fraction of $\kappa_{\rm unmagn}$,
while $\kappa_{\rm dynamic}$ is given by \Eq{dyn}.
Thus $\kappa_{\rm dynamic}/\kappa_{\rm particle}\sim \beta \alpha M_{\rm s} (L/\lambda)$,
i.e.\ the ratio does not depend on $M_{\rm A}$ (horizontal line in \Fig{space}).
When $M_{\rm A}>(L/\lambda)^{1/3}$ the mean free path of electrons is
constrained by $l_{\rm A}$.
In this case $\kappa_{\rm dynamic}/\kappa_{\rm particle}\sim \beta \alpha
M_{\rm s} M_{\rm A}^3$; see \Eqs{dyn}{el}.
This results in the separation line $\beta\alpha M_{\rm s} \sim M_{\rm A}^{-3}$
in \Fig{space}.

\begin{figure}\centering
\includegraphics[width=.8\textwidth]{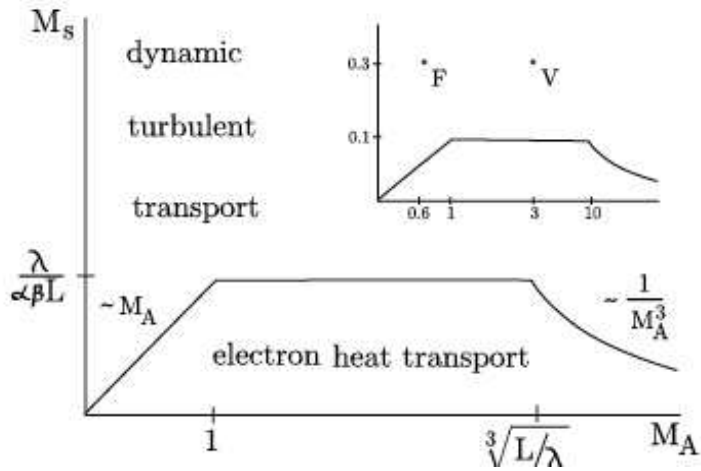}
\caption{Parameter space for particle diffusion or turbulent
diffusion to dominate: application to heat transfer. Sonic Mach
number $M_{\rm s}$ is plotted against the Alfv\'en Mach number $M_{\rm A}$. The heat
transport is dominated by the dynamics of turbulent eddies is above
the curve (area denoted ``dynamic turbulent transport'') and by thermal
conductivity of electrons is below the curve (area denoted ``electron heat
transport''). Here $\lambda$ is the mean free path of the electron, $L$ is
the driving scale, and $\alpha=(m_e/m_p)^{1/2}$, $\beta\approx 4$.
{\it Example of theory application}: The panel in the right upper corner of
the figure illustrates heat transport for the parameters for a cool core
Hydra cluster (point ``F''), ``V'' corresponds to the illustrative model
of a cluster core in \cite{EV06}.
Relevant parameters were used for $L$ and $\lambda$.
From L06.
}\label{space}\end{figure}

The application of the MHD approach to turbulent plasma has of course
its limitations.
For instance, in terms of magnetic reconnection, it is shown in
\cite{ELV11} that the model of turbulent reconnection described in LV99
is applicable to current sheets if the broadening of the current sheet
introduced through the wandering of magnetic field lines is larger than
the Larmor radius of thermal ions.
This makes the model not applicable to magnetosphere, where more
sophisticated, e.g.\ based on PIC simulations, modeling is required.

\subsection{Diffusion of magnetic fields in turbulent molecular clouds}

MHD turbulence induces not only mixing motions advecting heat, but it also
induces the transport of magnetic field and matter in molecular clouds.
This process, first discussed in \cite{Laz05} and \cite{LV09},
was later tested numerically in \cite{SantosLima10,SantosLima12}
and showed high efficiency for removing magnetic fields from clouds and
accretion disks.
\cite{LEC12} showed that the process that
they termed ``reconnection diffusion'' can explain why in observations by
\cite{Crutcher10} the envelopes had a lower mass to flux ratio than
the cloud cores. In contrast, the usually considered ambipolar diffusion
process predicts the opposite situation.

\begin{figure}
\includegraphics[width=0.95\columnwidth,height=0.24\textheight]{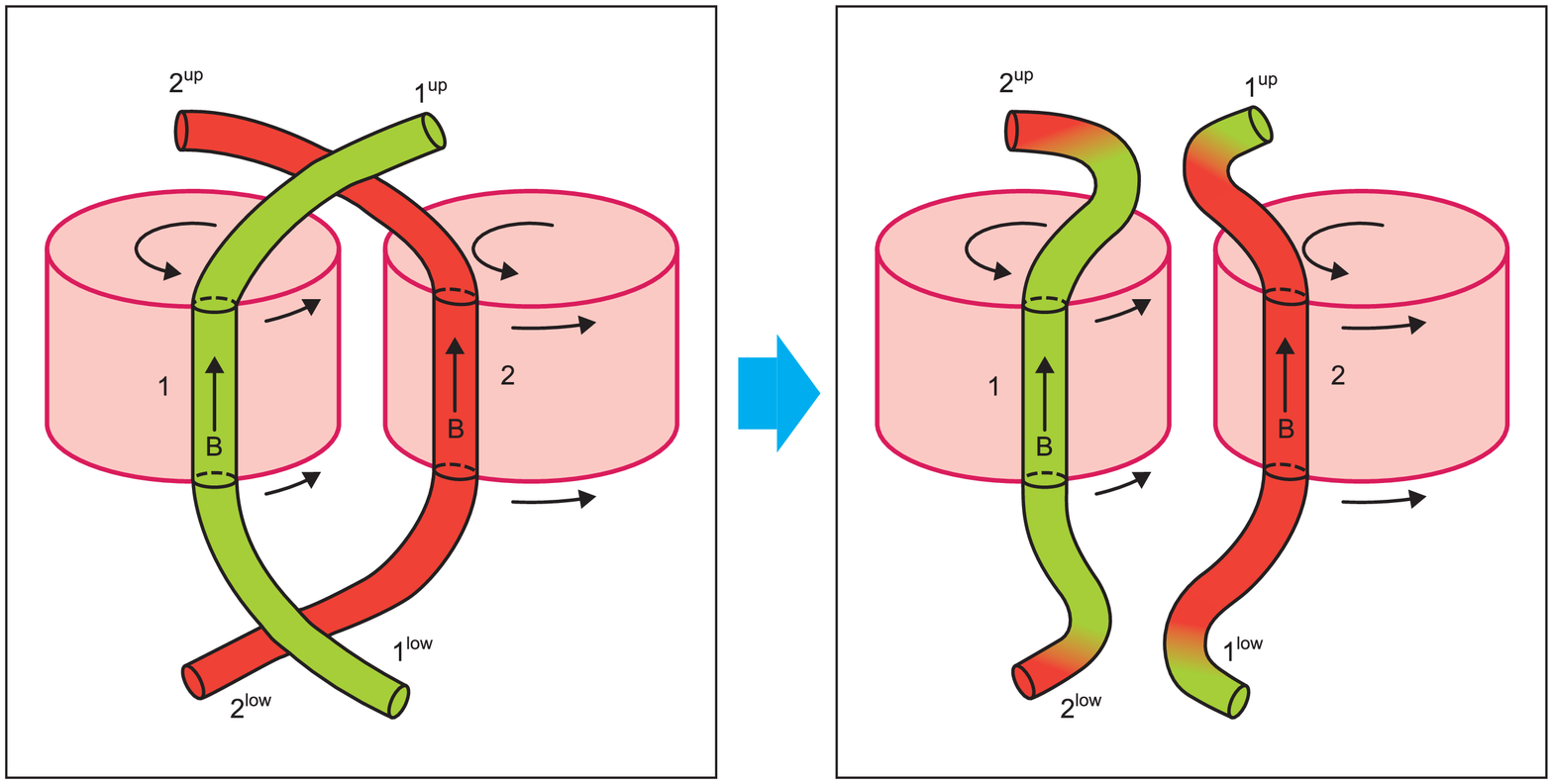}
\caption{Reconnection diffusion: exchange of flux with entrained
matter. Illustration of the mixing of matter and magnetic fields due
to reconnection as two flux tubes of different eddies interact. Only
one scale of turbulent motions is shown. In real turbulent cascade such
interactions proceed at every scale of turbulent motions. From \cite{Laz12}.
}\label{mix}\end{figure}

The elementary process of reconnection diffusion is illustrated in
\Fig{mix}, where the densities of plasma along magnetic flux
tubes belonging to different eddies are different.
The process of fast turbulent reconnection (LV99) creates new
flux tubes with columns of entrained dense and rarefied plasmas.
The situation is similar to the earlier discussed case with plasma
moving along magnetic fields and equalizing the pressure within the newly
formed flux tubes.
As a result, eddies with initially different plasma pressures exchange
matter and equalize the plasma pressure.
This process can be described as the diffusion of plasma perpendicular
to the mean magnetic field.
In reality, for turbulence with the extended inertial range, the
shredding of the columns of plasmas with different density proceeds at
all turbulence scales, making the speed of plasma motion irrelevant for the
diffusion.
For the case of strong turbulence, the diffusion of matter and magnetic
field is given by \Eq{dyn_weak}.
In the presence of the gravitational potential, the matter gets
concentrated towards the center of the potential well.
This was seen in the numerical simulations in \cite{SantosLima10}.
The physical justification of the process is based on the nature of
the GS95 cascade and the LV99 model of turbulent reconnection.
The deep relation between the two is discussed in \cite{ELV11}.

\subsection{Cosmic ray scattering, acceleration and perpendicular diffusion}

MHD turbulence plays an important role in accelerating energetic particles.
First of all, the second order Fermi acceleration can arise directly
from the scattering of particles by turbulence \citep{Mel68}.
Properties of MHD turbulence that we discussed above are essential to
understanding this process.
If turbulence is injected at large scales, the anisotropy of
Alfv\'enic modes at small scales makes them inefficient for scattering
and acceleration of cosmic rays \citep{Cha00,YL02}\footnote{The resonant
scattering is happening on the magnetic scales
of the order of the cosmic ray gyroradius.
If the Alfv\'enic eddies are strongly elongated, the particles
interacts with many eddies within its radius and the scattering effect
is dramatically reduced.
Scattering efficiency and the acceleration efficiencies are closely
related for the second order Fermi acceleration of cosmic rays by
turbulence (see Schlickeiser 2006).}.
In this situation, fast modes were identified in \cite{YL02}
as the major scattering and acceleration agent for cosmic rays and
energetic particles in interstellar medium \citep[see also][]{YL04,YL08}.
This conclusion was extended for solar environments in \cite{PYL06}
and intracluster medium in \cite{BL07}.
Turbulent magnetic field in the pre-shock and post-shock environment
are important for the first order Fermi acceleration associated with
shocks \citep{Sch02}.
In particular, magnetic field enhancement compared to its typical
interstellar values is important in the pre-shock region for the
acceleration of high energy particles.
The turbulent dynamo can provide a way of generating magnetic field in the
precursor of the shock.
In \cite{BJL09} it was shown that the interactions
of the density inhomogeneities pre-existing in the interstellar medium
with the precursor generate strong magnetic fields in the shock precursor,
which allows particle acceleration up to the energy of $10^{16}$\,eV.

While discussing heat transport by thermal electrons streaming
along turbulent magnetic fields, we have discussed the perpendicular
diffusion that is also relevant for the turbulent transport of cosmic
rays perpendicular to the mean magnetic field.
The relation between the parallel and perpendicular diffusivities in
this case is also given by \Eq{9}; see \cite{YL08}.
The important factor in this equation is $M_{\rm A}^4$.
This dependence follows from the modern theory of MHD turbulence and it
is very different from the dependence of $M_{\rm A}^2$ discussed in the
literature \citep{Jok74}.

A stream of cosmic ray protons propagating parallel or antiparallel to a large-scale
magnetic field can lead to important instabilities such as the Bell
instability \citep{Bell04}.
This is reviewed extensively in a companion paper by \cite{BBMO13}.
The combined presence of a cosmic ray current and a parallel magnetic
field gives rise to a pseudoscalar in the problem, and hence to
an $\alpha$ effect which can lead to large-scale dynamo action
\citep{RKBE12}.
In the following, we discuss magnetic field amplification by dynamo
action in more detail.

\section{MHD turbulence with dynamo-generated magnetic fields}

In this section we discuss the case where the magnetic field is produced
self-consistently by the action of turbulence through dynamo action.
We discuss here mainly the results of numerical simulations.

\subsection{Definitions and conventions}

In the following we characterize turbulent flows by the Reynolds
number, which quantifies the ratio of advective to viscous accelerations,
$\uu\cdot\nab\uu$ and $\nu\nabla^2\uu$, respectively.
Here, $\uu$ is the velocity and $\nu$ is the kinematic viscosity.
Throughout the remainder of this review, we define the Reynolds number as
\begin{equation}
\Rey=\urms/\nu\kf,
\label{Rey}
\end{equation}
where $\urms=\bra{\uu^2}^{1/2}$ is the rms velocity within some
appropriate volume and $\kf$ is the wavenumber of the energy-carrying
eddies, which is also known as the integral or correlation wavenumber.
It can be defined through a weighted average of the inverse wavenumber
over the kinetic energy spectrum, $\EK(k,t)$, where
$k=|\kk|$ is the modulus of the wave vector $\kk$, and $t$ is time.
The kinetic energy spectrum is normalized such that
\begin{equation}
\int_0^\infty\EK(k,t)\,\dd k=\half\rho_0\bra{\uu^2},
\end{equation}
where $\rho_0=\bra{\rho}$ is the volume average of the gas density $\rho$.
For incompressible and weakly compressible flows, it is customary to
ignore fluctuations of $\rho$ in the definition of $\EK(k,t)$.
In fact, there is no unique way of incorporating density.
For supersonic turbulence, this is very much a current research topic
in its own right.
We refer here to the papers of \cite{Kritsuk}, \cite{GB11}, and
\cite{BG13}.

Returning to the case of incompressible or weakly compressible (subsonic)
turbulence, a formal definition of $\kf$ can be written as
\begin{equation}
\kf^{-1}=\left.\int k^{-1} \EK(k,t)\,\dd k \right/ \int \EK(k,t)\,\dd k.
\label{kf}
\end{equation}
Note that $\kf=\kf(t)$ is in general time-dependent, which can be
important in studies of decaying turbulence.
An important example is helical MHD turbulence, because it drives an
inverse cascade which manifests itself in a time-dependent decrease
of $\kf(t)$; see \cite{TKBK12} and \cite{KTBN13} for recent examples.
In most of the cases considered below we consider a time average of $\kf$.

MHD turbulence is additionally characterized by the
{\it magnetic} Reynolds number,
\begin{equation}
\Rm=\urms/\eta\kf,
\label{Rey}
\end{equation}
where $\eta$ is the magnetic diffusivity.
The ratio $\Rm/\Rey=\nu/\eta=\Pm$ is the magnetic Prandtl number.
Furthermore, a magnetic energy spectrum $\EM(k,t)$ can be defined such that
\begin{equation}
\int_0^\infty\EM(k,t)\,\dd k=\half\mu_0^{-1}\bra{\BB^2},
\end{equation}
where $\BB$ is the magnetic field and $\mu_0$ is the vacuum permeability.
Analogously to \Eq{kf} we can then also define a magnetic correlation
wavenumber $\kM(t)$.
The relative alignment between $\uu$ and $\BB$ is characterized by the
so-called cross helicity, $\bra{\uu\cdot\BB}$, and its scale dependence
is characterized by the cross helicity spectrum $\EC(k,t)$ with the
normalization $\int\EC(k,t)\,\dd k=\bra{\uu\cdot\BB}$.
This quantity is a pseudoscalar and changes sign for a mirror-reflected
image of the turbulence.
Other important helicities are the kinetic helicity, $\bra{\ww\cdot\uu}$,
with $\ww=\nab\times\uu$ being the vorticity,
the current helicity, $\bra{\JJ\cdot\BB}$,
with $\JJ=\nab\times\BB/\mu_0$ being the current density,
and, in particular, the magnetic helicity, $\bra{\AAA\cdot\BB}$,
with $\AAA$ being the magnetic vector potential such that $\BB=\nab\times\AAA$.

In some cases we also discuss the evolution of a passive scalar,
whose concentration is governed by a corresponding diffusivity $\kappa$.
The relevant non-dimensional parameter is the P\'eclet number,
$\Pe=\urms/\kappa\kf$.

\subsection{Dynamo instability and spectrum}
\label{DynamoGenerated}

In the absence of an imposed magnetic field, the zero-field limit
is unstable to dynamo action when the magnetic Reynolds number
exceeds a critical value,
\begin{equation}
\Rm>\Rmc\quad\mbox{(dynamo instability)}.
\end{equation}
In practice, this means that the theory of Kolmogorov turbulence
is not directly applicable to most astrophysical flows when the gas
is ionized and therefore electrically conducting.

In this section we restrict ourselves to non-helical isotropic turbulence,
i.e., $\bra{\ww\cdot\uu}\ll\kf\bra{\uu^2}$.
In that case, only random or turbulent magnetic fields can be expected.
This possibility was already anticipated by \cite{Bat50}, but the relevant
theory was only developed later by \cite{Kaz68}.
He assumed that the velocity field was given by a smooth large-scale
random flow and found that the resulting magnetic field has typical
wavenumbers close to the resistive cutoff wavenumber,
$k_\eta=\bra{\mu_0^2\JJ^2/\eta^2}^{1/4}$, and much larger than $\kf$.
In fact, his work predicted a $k^{3/2}$ spectrum for the magnetic
field in the wavenumber range $\kf<k<k_\eta$.

The first numerical solutions of such dynamos have been performed by
\cite{MFP81} at a resolution of just $64^3$ collocation points.
Those where the ``golden years'' of numerical turbulence research.
For the first time, many of the ideas in turbulence could be put
to the test and, although the numerical resolution was still poor,
it was clear that it could only be a matter of time until all the
newly emerging results will be confirmed at better resolution.

In the following years, small-scale dynamo action emerged in several
direct numerical simulations (DNS).
At first it appeared that kinetic helicity had only a minor effect in
Cartesian simulations \citep{MP89,KYM91,Nor92}.
This was later understood to be an artefact of the lack of scale
separation, i.e., $\kf/k_1$ was not big enough \citep{HBD04}.
Meanwhile, global convection simulations in spherical shells did
produce large-scale magnetic fields \citep{Gil83,Gla85}.
Remarkably, although there was general awareness of the concepts of
large-scale and small-scale dynamos, which was also clearly spelled
out in an early review of \cite{VZ72}, the theory of \cite{Kaz68} was
still not yet widely cited in the West.
This has changed by the late 1990s \citep[e.g.,][]{GCS96,Sub98,Kul99},
and by the early 2000s many groups investigated the small-scale dynamo
systematically \citep{CV00,Scheko02,Scheko04,Schek04b,HBD03,HBD04}.

\begin{figure}[t!]\begin{center}
\includegraphics[width=.85\textwidth]{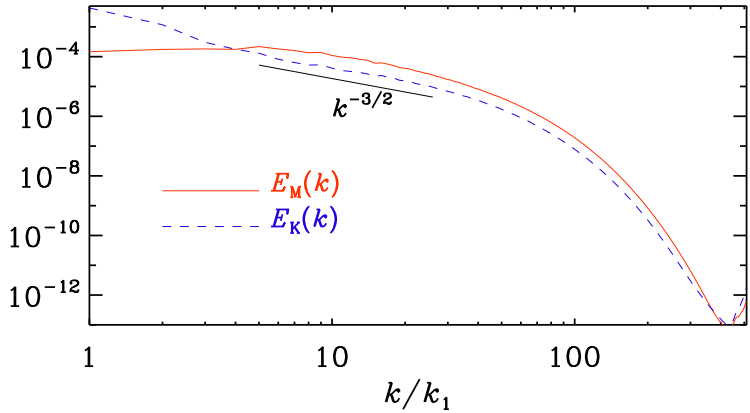}
\end{center}\caption[]{
Magnetic, kinetic and total energy spectra.
$\Rey=\Rm=960$ using $1024^3$ meshpoints.
Courtesy if Nils Erland Haugen \citep{HBD03}.
}\label{power1024a}\end{figure}

Although the resolution has improved significantly over the past
two decades, some important aspects of small-scale dynamos was evident
already early on.
In particular, \cite{MFP81} and \cite{KYM91} found that the magnetic
energy spectrum reaches a maximum at a wavenumber $k_{\rm M}$ that is
by a factor of $\approx6$ larger than $\kf$, which is where the kinetic
energy has its maximum.
This was an aspect that was later motivated by the work of \cite{Sub98},
who proposed that $k_{\rm M}/\kf$ should be of the order of $\Rmc^{1/2}$.
This result was indeed borne out by all the DNS obtained so far.
In \Fig{power1024a} we reproduce the result of \cite{HBD03} using
$1024^3$ meshpoints.
For larger values of $\Pm$, $\Rmc$ increases, so $k_{\rm M}$ also increases,
making it harder to confirm the expected scaling in that regime.
Indeed, \cite{Schek04b} propose that at large values of $\Pm$
the field shows folded structures.
While \cite{BS05} confirmed the presence of folded structures in a
simulation with $\Pm=50$, they found them rather the exception and
showed other cases where the field was not folded.
Recent simulations by \cite{BS13} confirmed that, after sufficiently
many turnover times, $k_{\rm M}/\kf$ is of the order of $\Rmc^{1/2}$
even when $\Pm=50$.

\begin{figure}[t!]\begin{center}
\includegraphics[width=.8\textwidth]{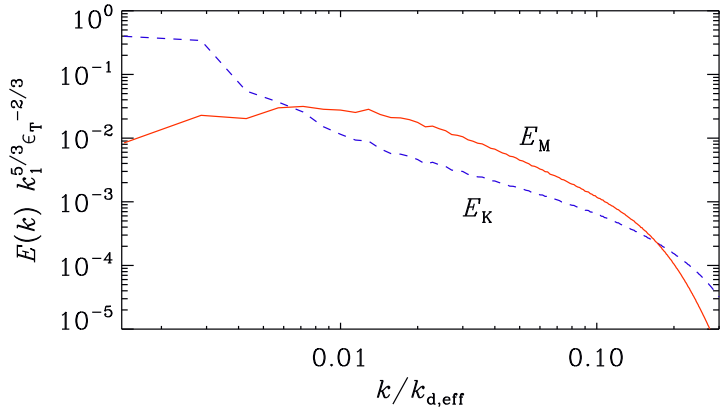}
\end{center}\caption[]{
Magnetic and kinetic energy spectra for runs with $512^3$
meshpoints and hyperviscosity with hyperresistivity (solid line) and
Smagorinsky viscosity with hyperresistivity (red, dashed line).
Note the mutual approach of kinetic and magnetic energy spectra
before entering the dissipative subrange.
Adapted from \cite{HB06}.
}\label{hyper_512}\end{figure}

Note that, at the position where the magnetic energy spectrum peaks,
the magnetic field is in super-equipartition with the kinetic energy
by a factor of 2--3.
Initially, this was a somewhat surprising result in view of the work of
GS95, according to which one might have expected equipartition.
Subsequent work using large eddy simulations suggested that this
super-equipartition would not persist deeper into the inertial range,
provided $\Rey$ and $\Rm$ are large enough.
Indeed, a trend toward equipartition can be seen in the compensated
energy spectra of \cite{HB06}; see also \Fig{hyper_512}, where
a Smagorinsky subgrid scale model was used for the momentum equation and
hyper-resistivity in the induction equation.

\subsection{Small-scale dynamo action at small values of $\Pm$}
\label{SmallScaleDynamoLowPm}

The question of what happens in the case of $\Pm\ll1$ has always been
on people's mind.
Small values of $\Pm$ are characteristic of denser bodies such as
stars, planets, and especially liquid metals.
Only in recent years a clearer picture has emerged of what happens
in the limit $\Pm\to0$.
By comparing the onset of dynamo action, it became clear that $\Rmc$
grew larger and larger as one approached the value $\Pm=0.1$ \citep{Scheko05}.
Crucial insight was gained through a paper by \cite{Iska}, who found that
$\Rmc$ has a local maximum at $\Pm=0.1$, and that it decreases again
as $\Pm$ is decreased further.
Early work of \cite{RK97} did already predict an increased value of $\Rmc$
in the limit of small values of $\Pm$, but not really a local maximum.
\cite{BC04} argue that the reason for an increased value of $\Rmc$ is
connected with the ``roughness'' of the velocity field, as quantified
by the scaling exponent $\zeta$ in velocity differences
$\delta u_\ell\sim\ell^\zeta$ over spatial separations $\ell$.
In the diffusive subrange, $\zeta=1$, so the velocity is smooth,
but in the inertial range we have $\zeta\approx0.4$, so velocity
gradients diverge and the velocity field is therefore called ``rough.''

The connection with roughness also helped explaining the occurrence
of a maximum in $\Rmc$ as $\Pm$ goes through 0.1.
Indeed, the reason for this is that near $\Pm=0.1$ the resistive wavenumber is
about 10 times smaller than the viscous one and thus right within the
``bottleneck'' where the spectrum is even shallower than in the rest
of the inertial range, with a local scaling exponent $\zeta\to0$,
corresponding to turbulence that is in this regime rougher still,
explaining thus the apparent divergence of $\Rmc$.

The physical reality of the bottleneck effect remains still a matter
of debate, but the work of \cite{Fal94} suggests that it is related
to the fact that near the viscous cutoff wavenumber the flow becomes
harder to stir, and that triangle interactions between a wavenumber
in the bottleneck range with wavenumbers in the dissipative subrange
experience a difficulty in disposing of their energy. 
It is claimed in \cite{BL10} that the MHD turbulence while
formally local, is more diffusive in terms of the interactions involved.
This property termed "diffuse locality", may explain that the bottleneck
effect in hydrodynamics is much more prominent that in MHD. Thus, one
may suspect that even the highest resolution simulations would still not be showing
the actual inertial range, but are influenced by an extended bottleneck effect
\cite{BL09}. This may be the reason why the numerically
measured spectrum is a bit shallower than the GS95 prediction. A numerical study
in Beresnyak (2011) seems to support this conclusion.

\begin{figure}[t!]\begin{center}
\includegraphics[width=.8\columnwidth]{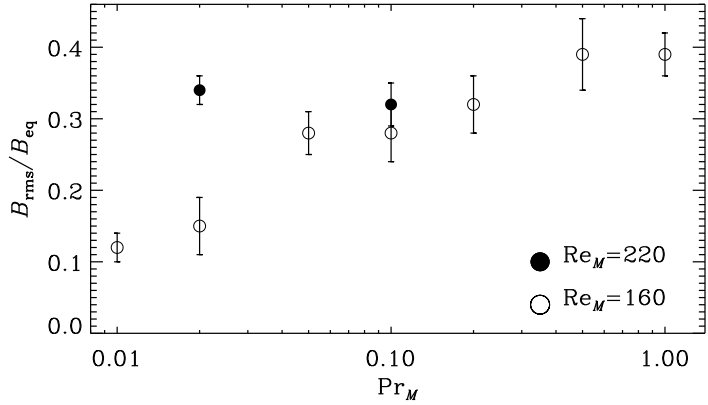}
\end{center}\caption[]{
Saturation field strengths for small-scale dynamos as a
function of $\Pm$ for two values of $\Rm$.
Note that for $\Rm=160$ (open symbols) the dynamo is close to onset and the
saturation field strength declines with decreasing values of $\Pm$,
while for $\Rm=220$ (filled symbols) the field strength changes only weakly
although only two data points are available.
}\label{pbsat}\end{figure}

It has recently become possible to demonstrate that in the nonlinear
regime, when the magnetic field affects the flow, the hydrodynamic bottleneck
effect tends to be suppressed as the field strength becomes appreciable,
so the divergence in the roughness
disappears and there is a smooth dependence of the saturation field
strength on the value of $\Pm$; see \cite{B11} for details.
In \Fig{pbsat} we show the saturation energy of small-scale dynamos
as a function of $\Pm$ using the data of Tables~1 and 2 of \cite{B11}.
It is clear that the position $\Pm=0.1$ is no longer special and
that dynamo action is possible for small values of $\Pm$ as well.
For $\Rm=160$ the value of $\Brms/\Beq$ is still $\Rm$-dependent,
but this may be an artefact of the dynamo being close to onset.
For $\Rm=220$ the dynamo is more clearly supercritical and, although
there are only two data points, the results are now more clearly
consistent with $\Brms/\Beq$ being independent of $\Rm$.

\begin{figure}[t!]\begin{center}
\includegraphics[width=.7\columnwidth]{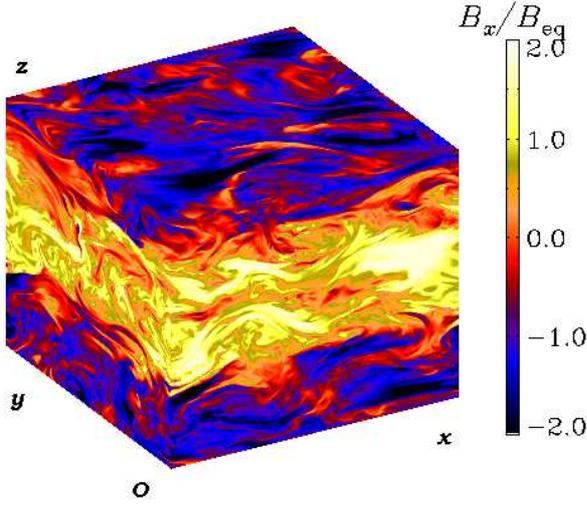}
\end{center}\caption[]{
Visualization of $B_x$ on the periphery of the computational
domain for a run with
$\Rm=600$ and a resolution of $512^3$ mesh points.
Note the clear anisotropy with structures elongated in the
direction of the field (which lies in the $xy$ plane).
Adapted from \cite{BRRS08}.
}\label{img_0138}
\end{figure}

\subsection{Helically driven turbulence}
\label{HelicallyDriven}

\begin{figure}[t!]\begin{center}
\includegraphics[width=\columnwidth]{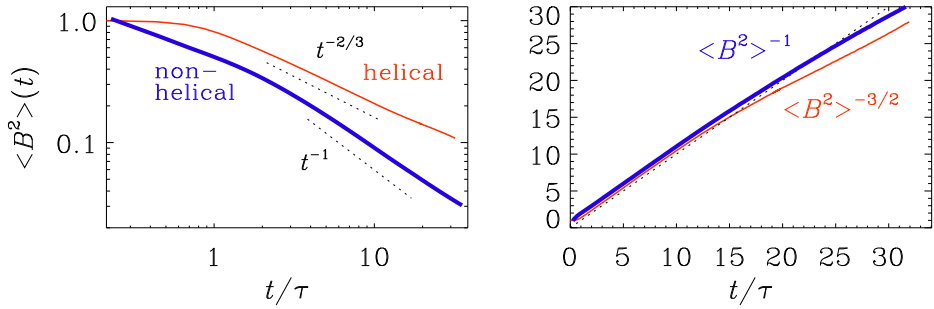}
\end{center}\caption[]{
Decay of magnetic energy with and without initial helicity (left)
and the approximately linear evolution of $\bra{\BB}^{-3/2}$ and
$\bra{\BB}^{-1}$ in the two cases (right).
}\label{pcomp_EM_EK_helnohel}
\end{figure}

Eigenfunctions of the curl operator provide an ideal means of
stirring the flow.
In wavenumber space, these take the form \citep{HBD04}
\begin{equation}
\ff_{\kk}=\RRRR\cdot\ff_{\kk}^{\rm(nohel)}\quad\mbox{with}\quad
{\sf R}_{ij}={\delta_{ij}-\ii\sigma\epsilon_{ijk}\hat{k}_k
\over\sqrt{1+\sigma^2}},
\label{Forcing}
\end{equation}
where $\sigma$ is a measure of the helicity of the forcing and
$\sigma=1$ for positive maximum helicity of the forcing function.
Furthermore,
\EQ
\ff_{\kk}^{\rm(nohel)}=
\left(\kk\times\eee\right)/\sqrt{\kk^2-(\kk\cdot\eee)^2}
\label{nohel_forcing}
\EN
is a non-helical forcing function, where $\eee$ is an arbitrary unit vector
not aligned with $\kk$; note that $|\ff_{\kk}|^2=1$ and
$\ff_{\kk}\cdot(\ii\kk\times\ff_{\kk})^*=2\sigma k/(1+\sigma^2)$,
so the relative helicity of the forcing function in real space is
$2\sigma/(1+\sigma^2)$.
When $\sigma=0$, the forcing function is non-helical, and so is
the resulting flow.
This case is special, as was demonstrated on various occasions.
Firstly, helical turbulence introduces an $\alpha$ effect which means
that a weak {\it large-scale} magnetic field becomes destabilized
and will be amplified.
In \Fig{img_0138} we show a visualization of one of the field
components on the periphery of a Cartesian domain with periodic
boundary conditions.
Note the presence of both large-scale and small-scale components.
Secondly, in the absence of forcing, a fully helical magnetic field
decays more slowly than a non-helical one.
Specifically, we have \citep{BM99,Bis03}
\EQ
\bra{\BB^2}(t)={\bra{\BB^2}(0)\over(1+t/\tau)^{2/3}},
\EN
where $\tau=\sqrt{\mu_0\rho_0}\bra{\AAA\cdot\BB}/\bra{\BB^2}^{3/2}$ is
the typical decay time scale.
In \Fig{pcomp_EM_EK_helnohel} we compare results of two simulations
of \cite{KTBN13}, one with an initial magnetic helicity and the other
one without.
Note the slower decay proportional to $t^{-2/3}$ in the helical case
compared to the faster $t^{-1}$ decay in the non-helical case.
In both cases, time has been normalized by $\tau=\sqrt{\mu_0\rho_0}/\kfz\Brms$,
where $\kfz=\kf(t=0)\approx15 k_1$.
The rms velocity is about 20\% of the $\Brms$ in the helical case
and about 28\% in the non-helical case.
The Reynolds number based on $\kf(t)$, which decreases with time
either like $t^{-2/3}$ in the helical case,
or like $t^{-1/2}$ in the non-helical case,
increases from 50 to 100 during the coarse of both simulations.
Even if the magnetic field is initially not fully helical, the relative
helicity will increase, because magnetic energy decays faster
than magnetic helicity; see \cite{TKBK12}.
These considerations are important for primordial magnetic fields
generated during cosmological phase transitions, because the inverse
cascade allow the fields to reach appreciable length scales at the
present time \citep{BEO96,BJ04,KTBN10}.

\begin{figure}[t!]
\centering\includegraphics[width=\columnwidth]{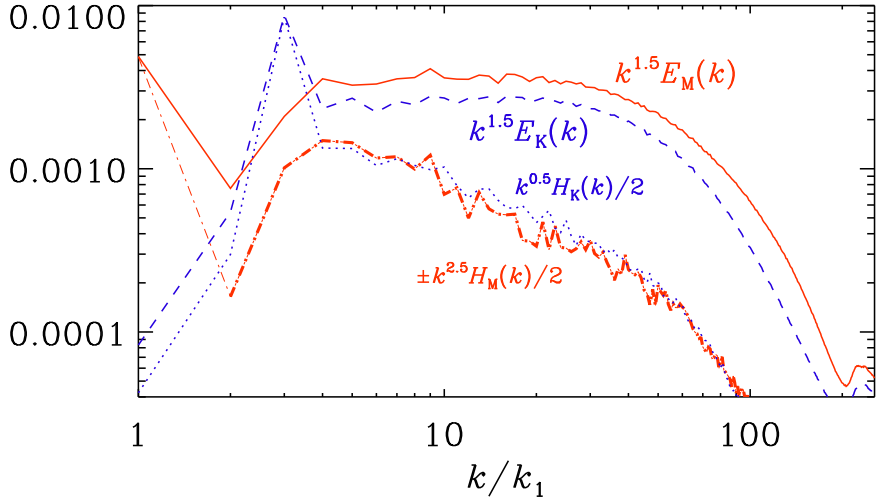}\caption{
Compensated time-averaged spectra of kinetic and magnetic energy
(dashed and solid lines, respectively), as well as of kinetic and
magnetic helicity (dotted and dash-dotted lines, respectively),
for a run with $\Rm\approx600$.
Note that $\HM(k)$ changes sign and becomes negative $k/k_1=1$
(indicated by a thin line), and is positive at all larger values
of $k/k_1$ (indicated by a thicker line).
Adapted from \cite{BRRS08}, where however $\HK(k)$ and $\HM(k)$
are compensated by $k^{1.2}$ and $k^{3.2}$, respectively.
}\label{pspec}\end{figure}

The $\alpha$ effect is the reason behind the large-scale dynamo effect
leading to the global magnetic field observed in many astrophysical
bodies \citep{Mof78,Par79,KR80}.
The resulting magnetic field is helical and its helicity has the
same sign as $\alpha$.
However, because of total magnetic helicity conservation,
no net magnetic helicity can be produced.
Therefore the $\alpha$ effect produces magnetic helicity of opposite
signs at large and small length scales at the same time.
In \Fig{pspec} we show magnetic and kinetic energy spectra compensated
by $k^{1.5}$ together with compensated magnetic and kinetic helicity spectra,
normalized by $k/2$ and $1/2k$, respectively.
This normalization allows us to see whether or not the realizability
conditions, $\EM(k)\ge2k\HM(k)$ and $\EK(k)\ge2\HK(k)/k$, are close to
being saturated.
Note also that $\HM(k)$ changes sign and becomes negative at $k/k_1=1$
(thin line), and is positive at all larger values of $k/k_1$ (thick line).

The case of homogeneous helical turbulence is a particularly interesting
example, because accurate estimates can be made about the saturation
field strength and the magnetic helicity balance, for example.
However, such circumstances are not usually found in realistic applications.
The significance of homogeneity is that then the divergence of the
magnetic helicity fluxes vanishes and does not affect the magnetic
helicity evolution, so we have
\begin{equation}
{\dd\over\dd t}\bra{\AAA\cdot\BB}=-2\eta\mu_0\bra{\JJ\cdot\BB}.
\label{ABm}
\end{equation}
Furthermore, in a homogeneous system, $\bra{\AAA\cdot\BB}$
is gauge-invariant, so in the steady state we have
\begin{equation}
\bra{\JJ\cdot\BB}=0\quad\mbox{(steady state)}.
\label{JBm}
\end{equation}
This is remarkable and applies even (and especially) in the case
of helical forcing when large-scale fields can be generated by
the $\alpha$ effect.

For the rest of this review, it will be crucial to distinguish
between large-scale and small-scale magnetic fields.
We do this by making use of the following decomposition:
\begin{equation}
\UU=\meanUU+\uu,\quad
\BB=\meanBB+\bb.
\end{equation}
In the previous sections of this review, there was no mean flow,
so $\UU=\uu$, but from now on we shall denote the full velocity
by a capital letter.
Likewise, the vorticity of $\UU$ is given by $\WW=\nab\times\UU$.

\begin{figure}[t!]\begin{center}
\includegraphics[width=.7\columnwidth]{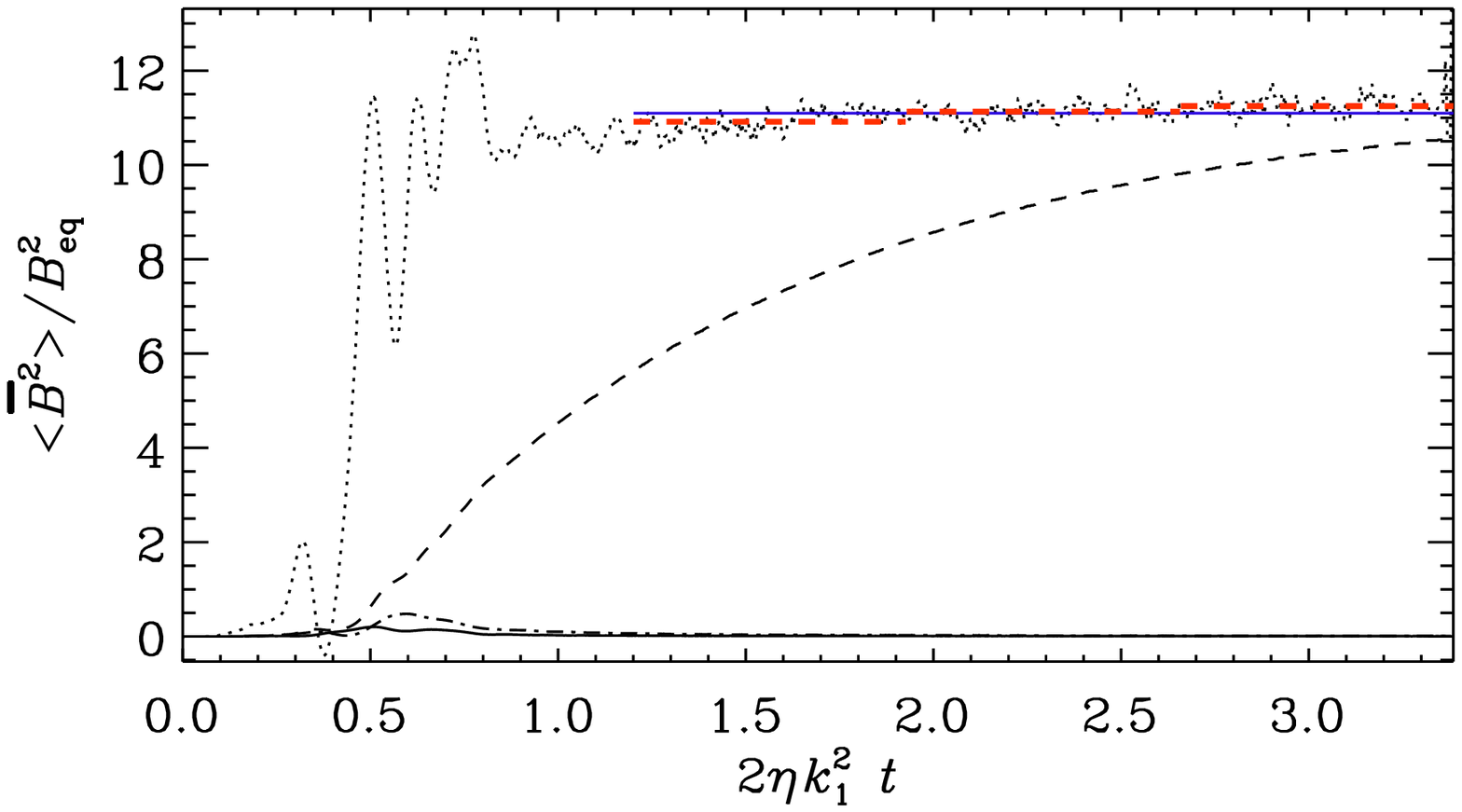}
\end{center}\caption[]{
Example showing the evolution of the normalized $\bra{\meanBB^2}$ (dashed)
and that of $\bra{\meanBB^2}+\dd\bra{\meanBB^2}/\dd(2\eta k^2t)$ (dotted),
compared with its average in the interval $1.2\leq2\eta k_1^2t\leq3.5$
(horizontal blue solid line), as well as averages over 3 subintervals
(horizontal red dashed lines).
Here, $\meanBB$ is evaluated as an $xz$ average, $\bra{\BB}_{xz}$.
For comparison we also show the other two averages, $\bra{\BB}_{xy}$
(solid) and $\bra{\BB}_{yz}$ (dash-dotted), but their values are very small.
Adapted from \cite{CB13}.
}\label{psat}\end{figure}

In rotating astrophysical bodies, a commonly used average is the
azimuthal one.
However, in the present case of fully periodic Cartesian domains,
the resulting large-scale fields can be described by planar averages,
such as $xy$, $yz$, or $xz$ averages.
The resulting mean fields, $\meanBB$, depend then still on
$z$, $x$, or $y$, in addition to $t$.
Examples of such fields are those proportional to $(\sin kz,\cos kz,0)$, 
$(0,\sin kx,\cos kx)$, and $(\cos ky,0,\sin ky)$, respectively.
All these examples obey
\begin{equation}
\nab\times\meanBB=k\meanBB
\end{equation}
and are thus eigenfunctions of the curl operator with eigenvalue $k$.
In particular, it follows then that $\meanJJ\cdot\meanBB=k\meanBB^2/\mu_0$
is uniform.
This can only be compatible with \Eq{JBm}, if there is a residual
(small-scale or fluctuating) magnetic field, $\bb=\BB-\meanBB$,
which obeys $\bra{\jj\cdot\bb}=-\bra{\meanJJ\cdot\meanBB}$.
Here, $\jj=\nab\times\bb/\mu_0$ is the corresponding current density.
Assuming $\bra{\jj\cdot\bb}=\epsf\kf\bra{\bb^2}/\mu_0$, we find that
$\meanBB^2/\bra{\bb^2}=\epsf\kf/k$, which can exceed unity in
cases of fully helical forcing ($\epsf\to\pm1$).
We recall that the parameter $\epsf$ is related to the helicity parameter
$\sigma$ in the forcing function \eq{Forcing} via $\epsf=2\sigma/(1+\sigma^2)$.
Repeating this calculation for the late saturation phase of
a dynamo, we have
\begin{equation}
{\meanBB^2\over\bra{\bb^2}}\approx{\epsf\kf\over k}
\left[1-e^{-2\eta k^2(t-t_{\rm sat})}\right],
\end{equation}
with a suitable integration constant $t_{\rm sat}$, having to do
with just properties of the initial field strength.
This equation describes the late ($t>t_{\rm sat}$), resistively dominated
saturation phase of a helically driven dynamo of $\alpha^2$ type.
By differentiating this equation again, we can find
that the final saturation field strength,
$\meanB_{\rm sat}=\meanB_{\rm rms}(t\to\infty)$, obeys \citep{CB13}
\begin{equation}
\meanB_{\rm sat}^2\approx\meanBB^2+\dd\meanBB^2/\dd(2\eta k^2t).
\end{equation}
This equation allows one to compute the value of $\meanB_{\rm sat}$
based on the measured rate at which $\meanBB^2$ increases.
It is now routinely used to estimate $\meanB_{\rm sat}$ without actually
reaching the the final state; see \Fig{psat} for an example.

\subsection{Turbulent mixing and non-diffusive transport}
\label{TurbulentMixing}

Turbulent flows are known to be capable of enhanced mixing.
A prime example is the mixing of a passive scalar concentration $C(\xx,t)$,
whose evolution is governed by the equation
\begin{equation}
{\partial C\over\partial t}=-\nab\cdot\left(\UU C\right)+\kappa\nabla^2C.
\end{equation}
Loosely speaking, turbulent mixing can be modeled as an enhanced
diffusivity in the corresponding evolution equation for the {\it mean}
passive scalar concentration $\meanC(\xx,t)$, which then takes the form
\begin{equation}
{\partial\meanC\over\partial t}=-\nab\cdot\left(\meanUU\,\meanC\right)
+\kappaT\nabla^2\meanC,
\label{dmeanCdt}
\end{equation}
where $\kappaT=\kappa+\kappat$ is the sum of molecular (or atomic)
and turbulent diffusivities.

In a more precise formulation, $\kappat$ becomes not only a tensor,
$\kappa_{ij}$, but also an integral kernel that takes into account that on
the right-hand side of \Eq{dmeanCdt} higher-order derivatives of $\meanC$
in space and time appear.
In particular, there can in principle also be a term of the form
$\nab\cdot(\ggamma^C\meanC)$ on the right-hand side which describes
turbulent pumping or turbophoresis, and $\ggamma^C$ is a vector.
This term acts like advection, but without any material motion.
(In a kernel formulation, such a term could in principle be subsumed
into the integral kernel.)
However, under isotropic conditions, $\ggamma^C$ must vanish and
the diffusivity tensor $\kappa_{ij}$ becomes an isotropic tensor
$\kappat\delta_{ij}$.
Analogous equations can also be derived for the magnetic induction
equation and the momentum equation.
In both cases this can lead to physically new effects such as the mean-field
(or large-scale) dynamo instability and the negative effective magnetic
pressure instability (NEMPI), which will be discussed further below.
The former exists in isotropic turbulence, while the latter requires
inhomogeneity and sufficiently strong density stratification.

In the simulations presented in \Sec{HelicallyDriven} we found the
development of large-scale fields of Beltrami type.
Such fields do indeed emerge as eigenfunctions of the related mean-field
induction equation with constant coefficients,
\begin{equation}
{\partial\meanBB\over\partial t}=\nab\times\left(\meanUU\times\meanBB
+\alpha\meanBB-\etaT\mu_0\meanJJ\right).
\label{dmeanBBdt}
\end{equation}
Significant progress in this field has recently become possible through
the numerical determination of the full set of turbulent transport
coefficients.
This method is known as the test-field method and involves the solution
of additional evolution equations for the magnetic fluctuations arising
from a given test field.
One needs enough test fields to obtain all tensor components.
By allowing the test fields to attain suitable variability in space
and time, it is possible to determine then also the full integral
kernel in spectral space.

The results obtained so far have shown that
\begin{equation}
\alpha\approx\alpha_0\equiv-\epsf\urms/3
\end{equation}
and
\begin{equation}
\etat\approx\etatz\equiv\urms/3\kf
\end{equation}
for $\Rm\ga1$.
For $\Rm\la1$, both coefficients increase linearly with increasing $\Rm$
\citep{SBS08}.
In the nonlinear regime, there can also be velocity fluctuations generated
through the presence of a mean field, but this requires what is known as
magnetic background turbulence, i.e., magnetic fluctuations that would
be present even without a mean magnetic field.
This is in principle possible when there is small-scale dynamo action.
This case can be treated with a correspondingly modified test-field
method \citep{RB10}.

In the following we state several important results obtained by using
the test-field method.
We did already mention that for fully helical turbulence and large values
of $\Rm$, $\alpha$ and $\etat$ attain values of the order of $\pm\urms/3$
and $\urms/3\kf$, respectively.
In turbulence, both coefficients possess a wavenumber dependence that
is of the form of a Lorentzian proportional to $(1+k^2/\kf^2)^{-1}$,
corresponding to an exponential integral kernel proportional to
$\exp[-(z-z')\kf]$; see \cite{BRS08}.
When the mean field is non-steady, the memory effect can become important
and this leads to a dependence of the form $(1-\ii\omega\tau)^{-1}$,
corresponding to a kernel proportional to $\exp(-|t-t'|/\tau)$ for $t'<t$,
and 0 otherwise.
Here, $\tau\approx(\urms\kf)^{-1}$ is the correlation time.

\begin{figure}[t!]
\centering\includegraphics[width=.8\columnwidth]{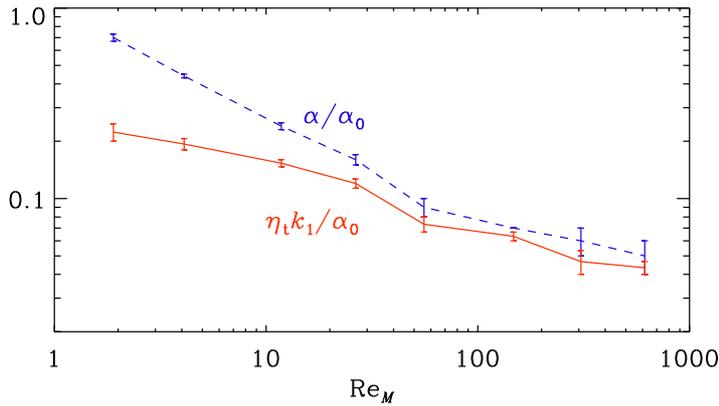}\caption{
$\Rm$-dependence of $\alpha$ and $\tilde{\eta}_{\rm t}$.
Both curves are normalized by $\alpha_0$.
Adapted from \cite{BRRS08}.
}\label{palpeta}\end{figure}

In the limit $\ell=1/\kf\to0$ and $\tau\to0$, the integral kernels
become $\delta$ functions in space and time.
However, this approximation breaks down at the bottom of the solar
convection zone, where the resulting mean magnetic field in dynamo models
often shows structures on scales much smaller than $\ell$ \citep{CGB11}.
Furthermore, nonlocality in time is violated when the mean magnetic field
is either growing or decaying.
Ignoring this can lead to discrepancies that are well detectable
with the test-field method \citep{HB09,Rad11}.
Finally, when the mean magnetic field depends strongly on both space and
time, the integral kernel in spectral space becomes approximately
proportional to $(1-\ii\omega\tau+\ell^2k^2)^{-1}$.
This form has the advantage that it can easily be treated in real space
by solving an evolution equation in time with a positive diffusion term, i.e.,
\EQ
\left(1+\tau{\partial\over\partial t}
-\ell^2\nabla^2\right)\meanemf_i
=\alpha_{ij}\meanB_j+\eta_{ijk}\meanB_{j,k}.
\label{emf_nonlocal}
\EN
Here, $\alpha_{ij}$ and $\eta_{ijk}$ are the usual $\alpha$ effect and
turbulent diffusivity tensors for $\omega\to0$ and $k\to0$, and equal
to $\alpha\delta_{ij}$ to $\etat\epsilon_{ijk}$ in the isotropic case.
This equation has been studied in some detail by \cite{RB12}.
It is a special form of the telegraph equation, which has been
studied in similar contexts \citep{BKM04,CSS13}.

Using the quasi-kinematic test-field method, \cite{BRRS08} showed that
in the case of a saturated dynamo, both $\alpha$ and $\etat$ remain
weakly $\Rm$-dependent; see \Fig{palpeta}.
Note that no fully asymptotic regime has been obtained yet, so it remains
unclear when or whether this will happen.
It is clear, however, that $\alpha$ must approach $\etat k_1$ at
large $\Rm$ for the system to be in the stationary saturated state.
However, in view of the astrophysical importance of turbulent dissipation,
the remaining weak dependence of $\etat$ on $\Rm$ is expected to disappear
eventually.

\section{Inhomogeneous MHD turbulence}

\subsection{Density stratification}

\begin{figure}
\includegraphics[width=.8\textwidth]{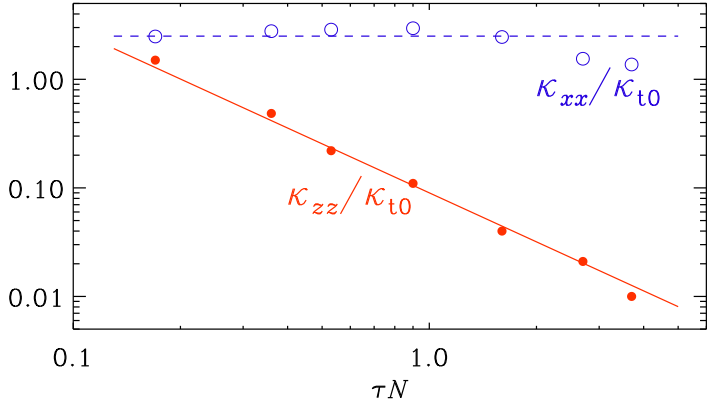}
\caption{Dependence of $\kappa_{xx}/\kappatz$ (open symbols) and
$\kappa_{zz}/\kappatz$ (filled symbols) on the normalized buoyancy frequency.
The dashed line shows that $\kappa_{xx}/\kappatz\approx2.5$ while the
solid line gives $\kappa_{zz}/\kappatz\approx0.09\,(\tau N)^{-3/2}$.
Adapted from \cite{KB12}.
}\label{pkap}
\end{figure}

Stratification refers to nonuniformity that is usually caused by gravity.
As a consequence, pressure increases in the direction of the gravity,
and this causes similar changes in density and/or temperature.
The turbulence intensity can itself also be stratified.
This usually comes as a consequence of density stratification,
but one can envisage circumstances in which the forcing is nonuniform.
Such non-uniformity affects turbulent transport--not just diffusive
but also non-diffusive transport, similar to the pumping velocity
proportional to $\ggamma^C$, described in \Sec{TurbulentMixing}.
Both effects are astrophysically important.
Stratification usually leads to a suppression of diffusive transport.
An example is shown in \Fig{pkap}, where we show the suppression of
the vertical passive scalar diffusivity as a function of the
stratification, which is here measured by the normalized
Brunt--V\"ais\"al\"a frequency, $N$, with $N^2=-\grav\cdot\nab s/c_p$
and $s$ being the specific entropy.
For details of this, see the work of \cite{KB12}.

Suppression of turbulent transport, for example, is critical
for understanding the depletion of primordial elements (e.g., lithium)
by mixing with deeper layers in the stably stratified lower overshoot
layer of the convection zones of stars with outer convection zones.
The suppression is here caused mainly by the stabilizing entropy gradient
[reversing the gradient of $s$ leads the negative values of $N$,
corresponding the onset of convection with exponential growth
proportional to $\exp(\mbox{Im} N t)$.]
In the following, we shall focus on another manifestation of stratification,
namely the expansion of rising structures as they ascent into less
dense surroundings.
For that purpose, we make the assumption of an isothermal equation of
state, which is a simplification that leads to a constant pressure scale
height and suppresses also the stabilizing effect from the entropy gradient.
For further discussion on this, see the papers by \cite{BKKR12} and
\cite{KBKMR12} in the context of NEMPI; see \Sec{TurbulentMixing}.

\subsection{Stratified turbulence with a vertical field}

\begin{figure}[t!]
\begin{center}
\includegraphics[width=.8\columnwidth]{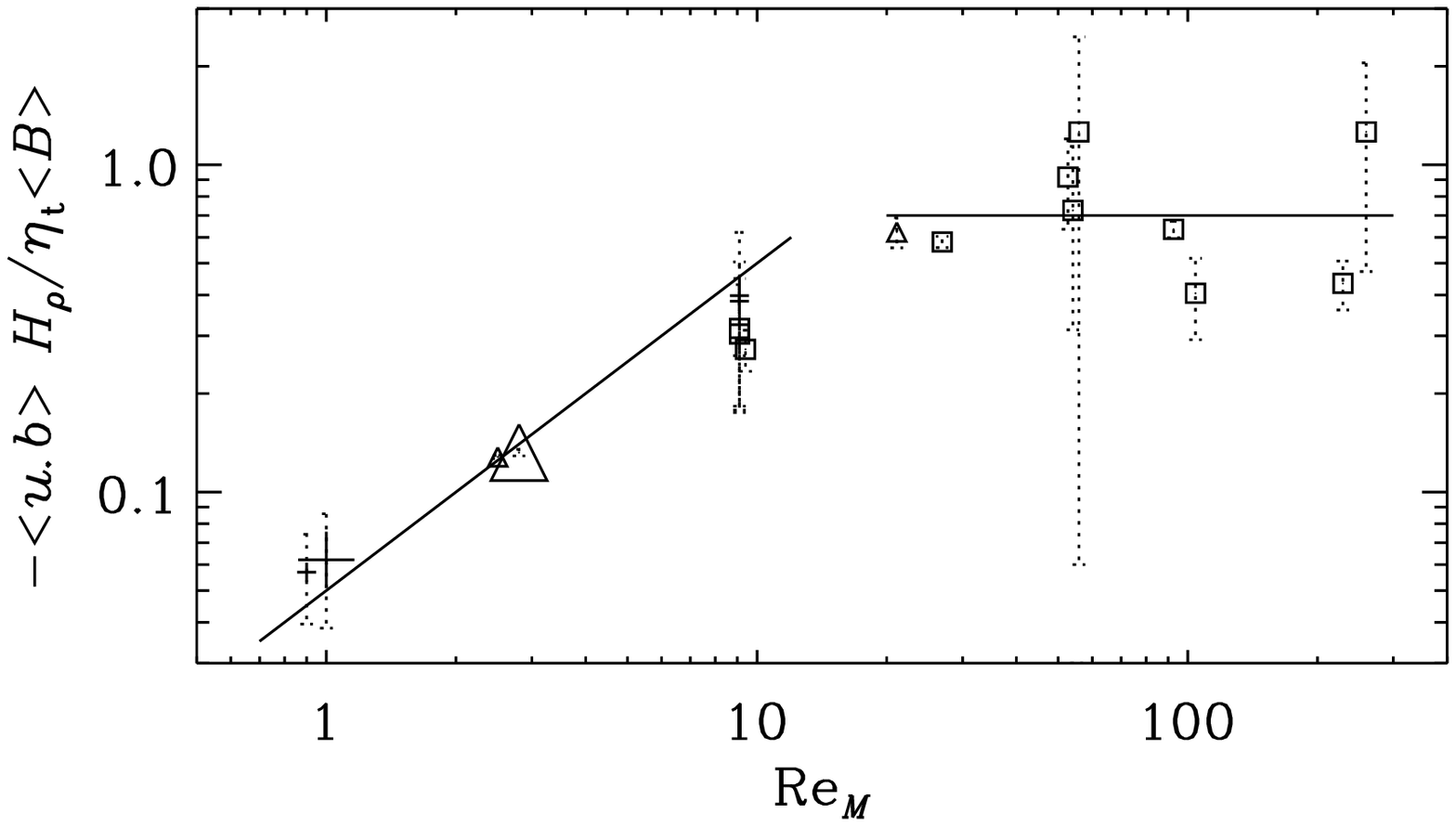}
\end{center}
\caption[]{
Dependence of the normalized cross helicity on ${\rm Rm}$ for
various field strength $B_z/B_{\rm eq}<0.1$, ${\rm Pm}=1$, $k_{\rm
f}/k_1=2.2$, and $H_\rho k_1=2.5$.
The straight line denotes the fit
$\langle\uu\cdot\bb\rangle/\tau\grav\bra{\BB}=0.05\,{\rm Rm}$.
}\label{pRm_dep}
\end{figure}

In the presence of stratification and an imposed magnetic field along
the direction of stratification, there is the possibility of
producing another pseudoscalar called cross helicity.
On theoretical grounds, one expects \citep{RKB11}
\begin{equation}
\bra{\uu\cdot\bb}\propto\grav\meanBB.
\end{equation}
More specifically, it turns out that
\begin{equation}
\bra{\uu\cdot\bb}=-\etat\meanB/H_\rho,
\end{equation}
where $H_\rho$ is the density scale height.
This does indeed turn out to be the case, as has been shown using
simulations of forced isothermal turbulence in the presence of gravity.
The result is shown in \Fig{pRm_dep}, where we plot $\bra{\uu\cdot\bb}$
as a function of $\Rm$.

\subsection{Effects of rotation}

In the presence of stratification and/or rotation, MHD turbulence is
subject to a range of new effects.
These phenomena are associated with the vectors $\grav$ (gravity) and
$\OO$ (angular velocity), which introduce preferred directions to the flow.
They do so in different ways, because $\grav$ is a polar vector
(identical to its mirror image) while $\OO$ is an axial vector
(antiparallel to its mirror image).
This means that turbulent transport effects characterized by some
effective velocity must be proportional to another polar vector.
This can then either be the vector $\grav$ or,
in forced turbulence simulations, where it is possible to produce
helical turbulence, it can be the vector $\OOO$.
In that case there is kinetic helicity, $\bra{\ww\cdot\uu}$,
which is a pseudoscalar, so $\bra{\ww\cdot\uu}\OO$
would also be a polar vector, allowing pumping even in the
homogeneous case if there is rotation and helicity.

\begin{figure}[t!]\begin{center}
\includegraphics[width=\columnwidth]{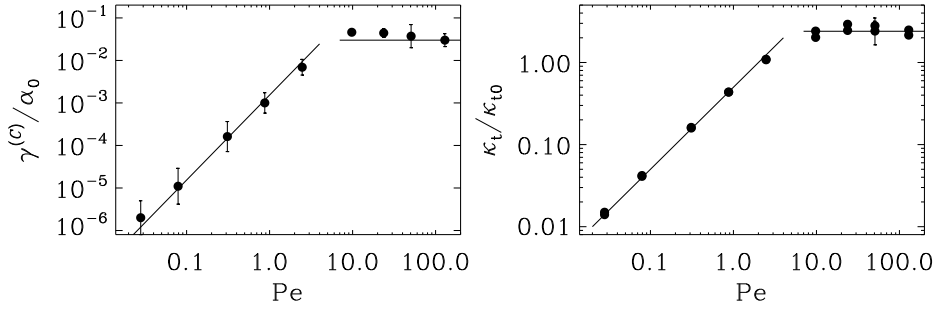}
\end{center}\caption[]{
Dependence of passive scalar pumping velocity, $\ggamma^{(C)}$,
and passive scalar diffusivity $\kappat$ on Peclet number, $\Pe$.
The scale separation ratio is $\kf/k_1=5$.
}\label{ppgam}\end{figure}

In \Fig{ppgam} we show such an example, where there is fully helical
turbulence that is initially isotropic, but because of rotation it
becomes anisotropic and there is now a polar vector that leads to
turbulent pumping with the velocity
\EQ
\ggamma^{(C)}\approx0.075\bra{\ww\cdot\uu}\OO/(\urms\kf)^2.
\EN
A similar result has previously been obtained by \cite{Pip08}
and \cite{MKTB09} for shear flows, where the resulting mean vorticity
vector acts as the relevant pseudovector.
However, these situations are somewhat artificial, because helicity does
not normally occur in the absence of additional stratification, so any
pumping would still be indirectly associated with the stratification vector,
although it can now attain a direction proportional to $\OOO$ or the
mean vorticity.

\begin{figure}[t!]\begin{center}
\includegraphics[width=\columnwidth]{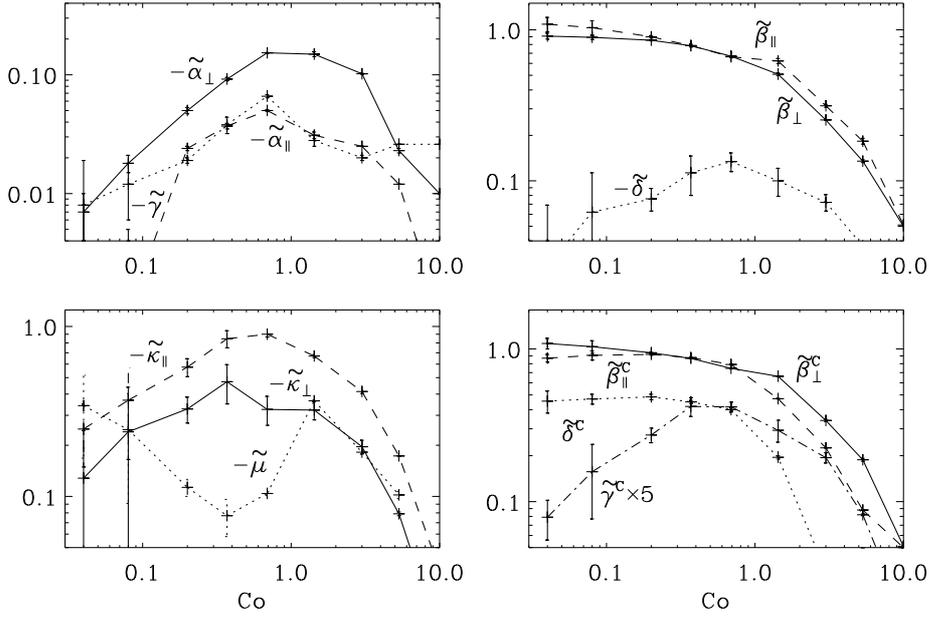}
\end{center}\caption[]{
Dependence of transport coefficients in a model with rotation and
density stratification as a function of the Coriolis number,
$\Co=2\Omega/\urms\kf$.
The other relevant parameters are $\Rm\approx10$,
$\mbox{Gr}=g/\cs^2\kf\approx0.16$, $\kf/k_1=5$, for $\nu=\eta=\kappa$.
}\label{psummary_Odep_g08}\end{figure}

Owing to the presence of stratification and rotation, the turbulence
attains helicity and can then produce an $\alpha$ effect.
This has been studied in great detail in the past using analytic methods
and, more recently, the test-field method.
In \Fig{psummary_Odep_g08} we show an example from \cite{BRK12},
where the turbulence is governed by only one preferred direction,
and $\OOO$ and $\grav$ are therefore assumed to be parallel.
In that case, $\meanEMF$ can be represented in the form
\EQA
\meanEMF&=&
-\alpha_\perp\meanBB
-(\alpha_\parallel-\alpha_\perp)(\eee\cdot\meanBB)\eee
-\gamma\eee\times\meanBB
\nonumber \\
&&-\beta_\perp\mu_0\meanJJ
-(\beta_\parallel-\beta_\perp)(\eee\cdot\mu_0\meanJJ)\eee
-\delta\eee\times\mu_0\meanJJ
\label{eq005}\\
&&-\kappa_\perp\meanKK
-(\kappa_\parallel-\kappa_\perp)(\eee\cdot\meanKK)\eee
-\mu\eee\times\meanKK
\nonumber
\ENA
with nine coefficients $\alpha_\perp$, $\alpha_\parallel$, $\ldots$, $\mu$.

Clearly, because of stratification and rotation, the turbulence is no
longer isotropic, so $\alpha$ will also no longer be isotropic.
In the simplest case when both $\grav$ and $\OO$ are parallel,
$\alpha$ has components parallel and perpendicular to their direction.
The $\alpha$ effect is of particular interest, because it can lead to
large-scale magnetic field generation.
Another effect that is known to lead to large-scale dynamo action is the
R\"adler or $\OO\times\JJ$ effect \citep{Rad69}.
Unlike the $\alpha$ effect, it exists already with just rotation
and no stratification.
Its astrophysical relevance is however still to be demonstrated.
Note also that in all practical situations there must still be an
additional source of energy, because $\OO\times\meanJJ$ has no component
along $\meanJJ$ and does therefore not provide energy to the system.

\subsection{Stratified turbulence with an imposed magnetic field}
\label{NEMPI}

In the presence of an imposed magnetic field there is an important
effect that deserves to be mentioned.
In mean-field parameterizations of the Reynolds stress, there are terms
that are quadratic in the mean magnetic field and contribute to a
decrease of the Reynolds stress if there is a weak magnetic field.
This suppression was discussed by \cite{Rue74} and \cite{Rae74} in
connection with the understanding of the quenching of the $\alpha$
effect by a mean magnetic field.
However, later it was understood that it also leads to a suppression
of the turbulent pressure and that this suppression is stronger
than the explicitly added magnetic pressure from the mean field,
$\meanBB^2/2\mu_0$.
This means that the contribution of the mean field
to the total turbulent pressure,
\begin{equation}
p_{\rm tot}=p_{\rm gas}+p_{\rm turb}=
p_{\rm gas}+p_{\rm turb}^{(0)}+
\left[1-\qp(\meanBB^2/\Beq^2)\right]\meanBB^2/2\mu_0,
\end{equation}
which is embodied by the last term,
$\left[1-\qp(\meanBB^2/\Beq^2)\right]\meanBB^2/2\mu_0$,
can be negative \citep{KRR89,KRR90,KMR93,KMR96,KR94,RK07}.
Here, $\qp(\meanBB^2/\Beq^2)$ is a non-dimensional quenching function
describing the suppression of the total stress,
which consists of Reynolds and Maxwell stress.
It is only a function of $\meanBB^2/\Beq^2$, so even for a uniform
$\meanBB^2$ it can show spatial variation if $\Beq^2$ changes, for example
as a result of density stratification.
This allows the full dependence of $\qp$ on $\meanBB^2/\Beq^2$ to be
probed in a single simulation \citep{BKKR12,KeBKMR12}.
The effect of the Maxwell stress turns out the be weaker than that of the
Reynolds stress and it has the opposite effect, as was demonstrated by
numerical calculations \citep{BKR10}.

\begin{figure}\begin{center}
\includegraphics[width=\textwidth]{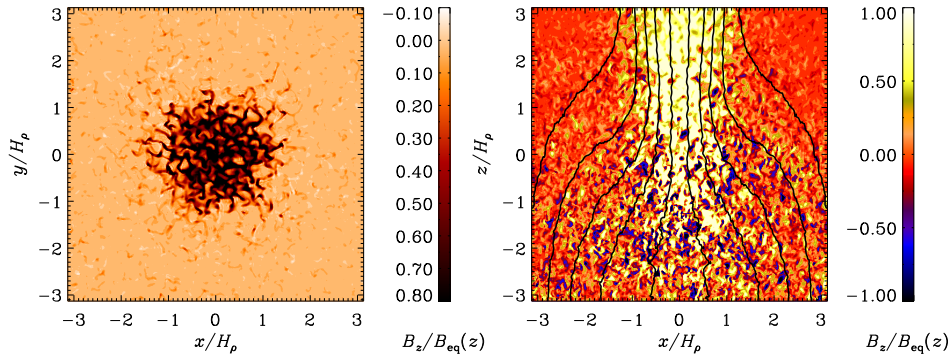}
\end{center}\caption[]{
Cuts of the vertical magnetic field in units of the equipartition field
strength, $\meanB_z/\Beq(z)$, through the horizontal plane at the top
boundary (left) and the vertical plane through the middle of the spot (right).
Field lines of the numerically averaged mean field are superimposed.
Adapted from \cite{BKR13}.
}\label{pslices_V256k30VF_Bz002}\end{figure}

In a stratified layer with a sub-equipartition magnetic fields
this negative effective magnetic pressure can lead
to an instability producing spontaneously magnetic flux concentrations
\citep{KRR89,KMR93,RK07}.
This has recently been confirmed with DNS \citep{BKKMR11,KeBKMR12}
and is being discussed in connection with explaining the spontaneous
formation of active regions \citep{KBKMR13} and sunspots \citep{BKR13}.
In \Fig{pslices_V256k30VF_Bz002} we show horizontal and vertical cuts
through a magnetic spot from the simulation of \cite{BKR13} in the
presence of an imposed vertical field.
In the horizontal cut, again, strong fields correspond to dark shades.
The vertical cut is with a different color table where strong fields
now correspond to light shades.
It shows that the magnetic field (in units of the local
equipartition field strength) decreases with height.
Note also that the mean magnetic field fans out toward
the bottom of the domain.
Applying this finding to the origin of sunspots, it suggest that,
contrary to common belief \citep[cf.][]{B05}, those structures
may not be deeply anchored.

\subsection{Solar dynamo and magnetic helicity fluxes}

One of the main applications of mean-field theory has always been
to explain the Sun's global magnetic field, its 11 year cycle,
and the migration of magnetic field from mid to low latitudes,
in addition, of course, eventually the formation of sunspots themselves.
In the last few years, several groups have engaged in tackling the
problem of the Sun's global magnetic field
by performing numerical simulations of rotating turbulent
convection in spherical shells using either spherical harmonics
\citep{MT09,BBBMT10,BMBBT11}, an implicit solver \citep{GCS10,RCGBS11},
or finite differences in spherical wedges \citep{KKBMT10,KMB12}
to overcome the timestep constraint at the poles.
The results from all groups trying to model the Sun agree in that they show
equipartition-strength magnetic fields in the bulk of the convection zone
(rather than highly super-equipartition-strength magnetic fields just at
the bottom of the convection zone), with magnetic activity concentrated
toward low latitudes and, in some cases, cyclic reversals of the magnetic
field direction, resembling the solar 22 year cycle.

\begin{figure}[t!]\begin{center}
\includegraphics[width=0.58\textwidth]{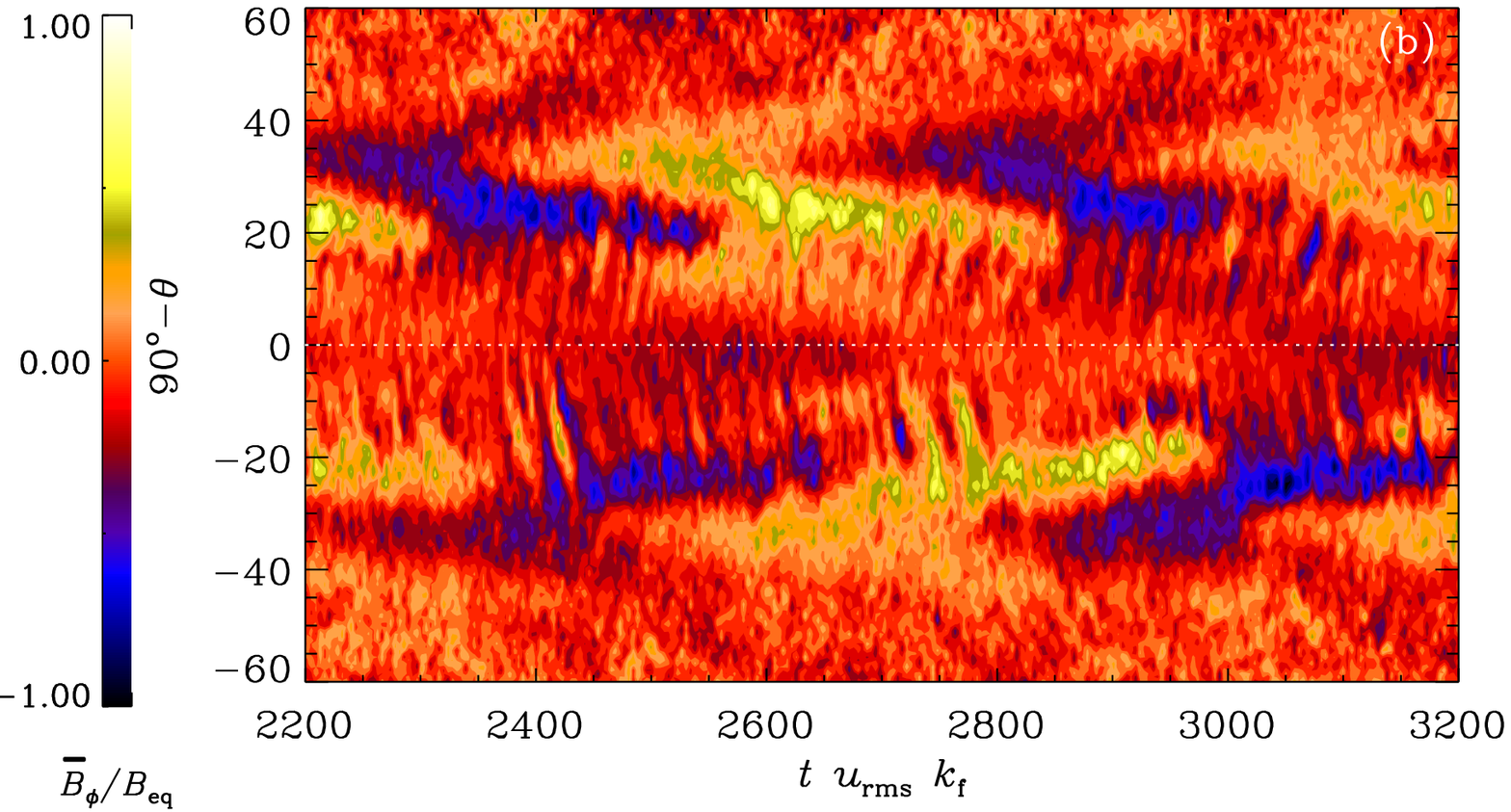}
\includegraphics[width=0.41\textwidth]{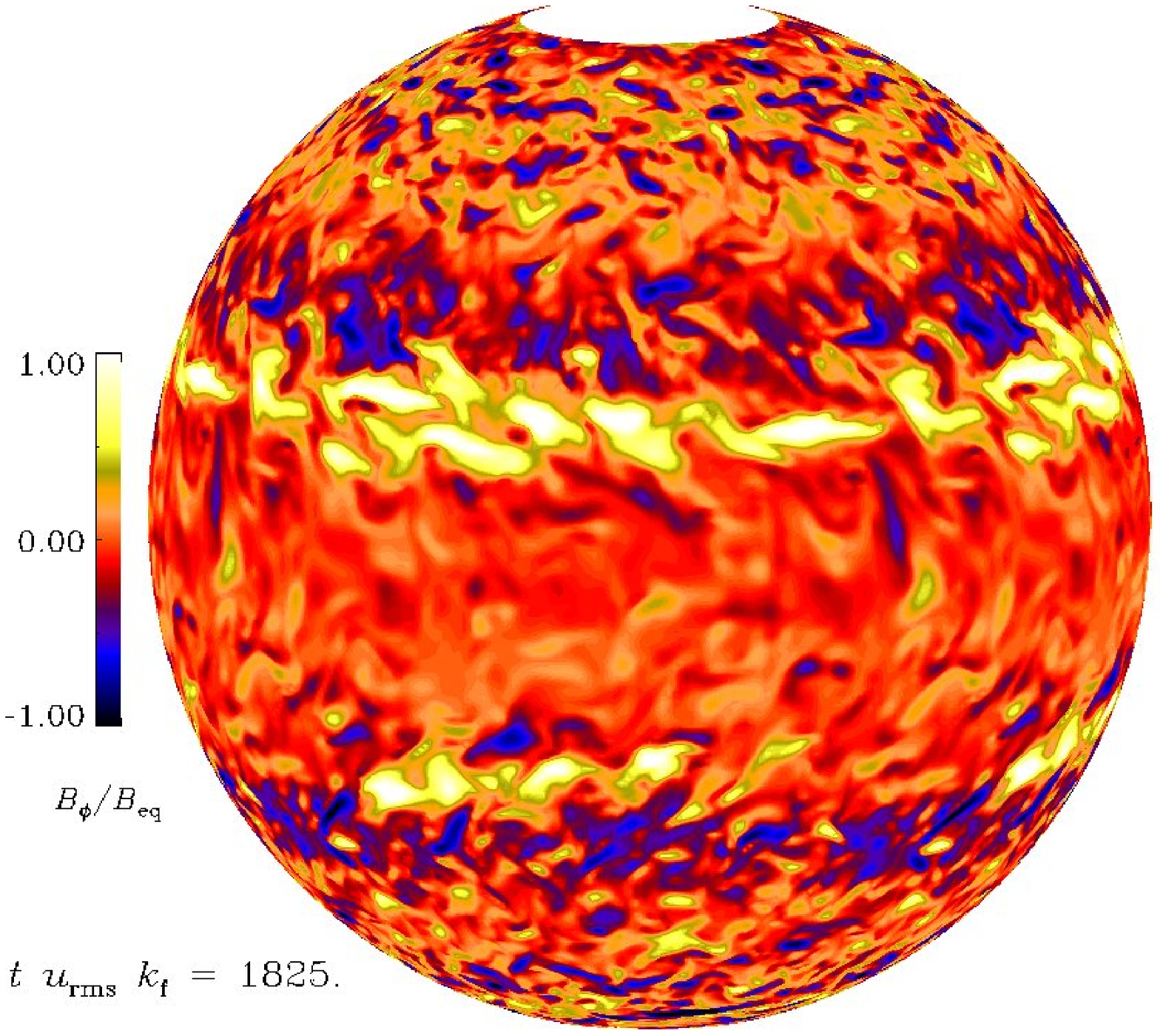}
\end{center}\caption{{\em Left}: azimuthally averaged toroidal magnetic field
as a function of time (in turnover times) and latitude
(clipped between $\pm60^\circ$).
Note that on both sides of the equator ($90^\circ-\theta=\pm25^\circ$),
positive (yellow) and negative (blue) magnetic fields move equatorward,
but the northern and southern hemispheres are slightly phase shifted
relative to each other.
{\em Right}: Snapshot of the toroidal magnetic field $B_\phi$ at $r=0.98$
outer radii.
Courtesy of \cite{KMB12}.
}{ }\label{equatorward}\end{figure}

A major breakthrough has been achieved through the recent finding of
equatorward migration of magnetic activity belts
in the course of the cycle \citep{KMB12}; see \Fig{equatorward}.
These results are robust and have now been reproduced in extended simulations
that include a simplified model of an outer corona \citep{WKMB12}.
Interestingly, the convection simulations of all three groups produce cycles
only at rotation speeds that exceed those of the present Sun by a factor of
3--5 \citep{BMBBT11}.
Both lower and higher rotation speeds give, for example, different
directions of the dynamo wave \citep{KMB12}.
Different rotation speeds correspond to different stellar ages
(from 0.5 to 8 gigayears for rotation periods from 10 to 40 days),
because magnetically active
stars all have a wind and are subject to magnetic braking \citep{Sku72}.
In addition, all simulations are subject to systematic ``errors'' in
that they poorly represent the small scales and emulate in that way an
effective turbulent viscosity and magnetic diffusivity that is larger
than in reality; see the corresponding discussion in Sect.~4.3.2 of
\cite{BKKR12} in another context.
In future simulations, it will therefore be essential to explore the range
of possibilities by including stellar age as an additional dimension of the
parameter space.

\begin{figure}[t!]\begin{center}
\includegraphics[width=.8\textwidth]{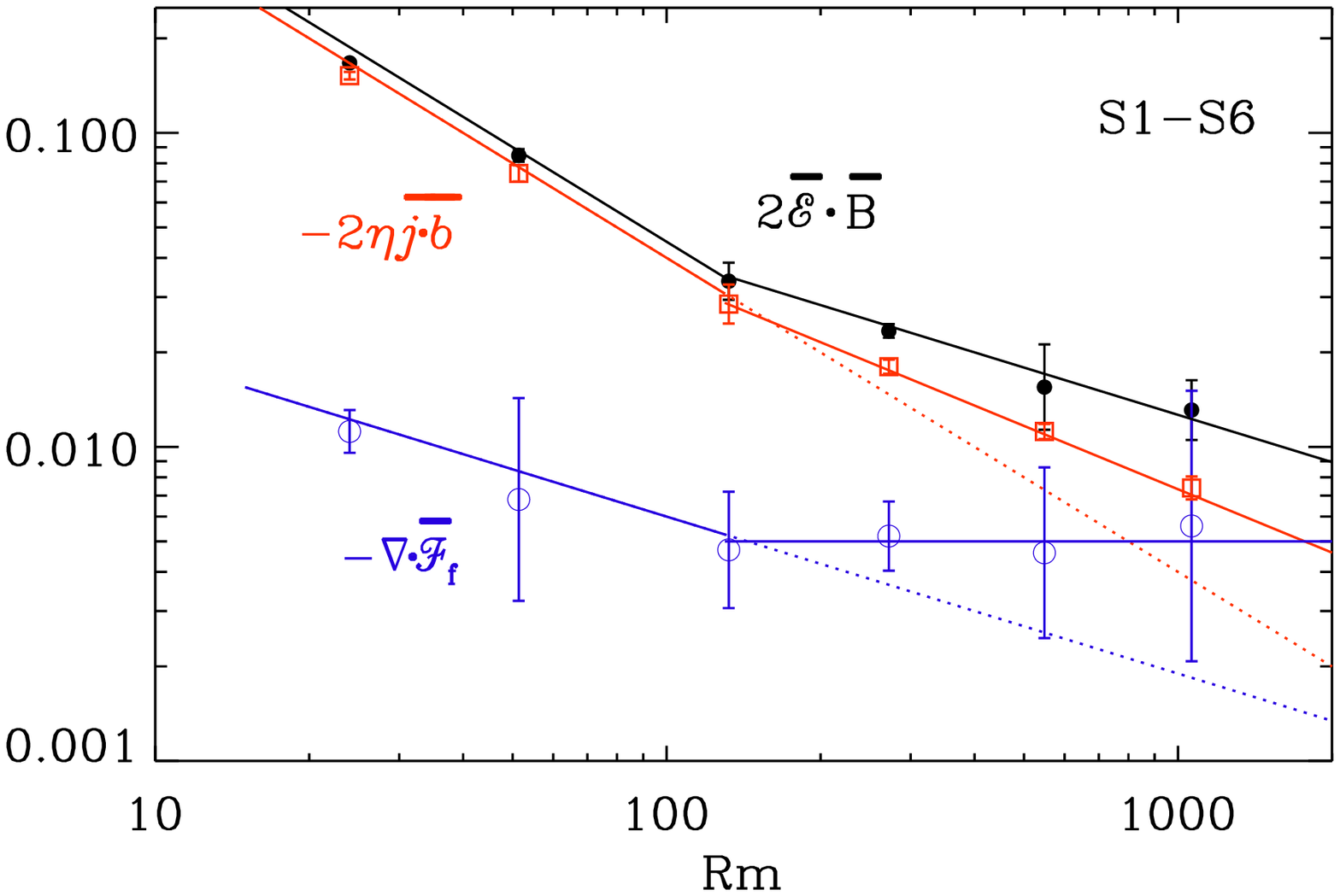}
\end{center}\caption[]{
Scaling properties of the vertical slopes of $2\meanEMF\cdot\meanBB$,
$-2\eta\mu_0\,\overline{\jj\cdot\bb}$, and $-\nab\cdot\meanFFf$.
The three quantities vary approximately linearly with $z$, so
the three labels indicate their non-dimensional values at $k_1z=1$.
The dotted lines show the extrapolated initial scaling for low $\Rm$.
Adapted from \cite{DSGB13}.
}\label{presults2b}\end{figure}

In support of our statement that a poor representation of the small
scales in DNS emulates artificially enhanced turbulent viscosity and
turbulent magnetic diffusivity, let us recall that $\etat$ and $\alpha$
are scale-dependent.
As discussed before in \Sec{TurbulentMixing}, they decrease with
increasing $k$ in a Lorentzian fashion.
The relative importance of $\Omega$ effect over the $\alpha$ effect
depends on the ratio of $C_\Omega$ and a similar parameter
$C_\alpha=\alpha/\etat k$ characterizing the strength of the $\alpha$ effect.
Both $C_\Omega$ and the ratio $C_\Omega/C_\alpha$ would be underestimated
in a large eddy simulation in which $\etat(k) k$ and $\alpha(k)$ are too
big, so one would need to compensate for this shortcoming by increasing
$\Omega$ to recover cyclic dynamo action.

As alluded to in \Sec{HelicallyDriven}, magnetic helicity fluxes
play a major role in the dynamo by alleviating the otherwise
catastrophic quenching of the dynamo \citep{BB03}.
Recent work using a simple model with a galactic wind has shown,
for the first time, that this may indeed be possible.
We recall that the evolution equation for the mean magnetic helicity density
of fluctuating magnetic fields, $\meanhf=\overline{\aaaa\cdot\bb}$, is
\begin{equation} {\partial\meanhf\over\partial t}=
-2\meanEMF\cdot\meanBB-2\eta\mu_0\,\overline{\jj\cdot\bb}
-\nab\cdot\meanFFFFf,
\end{equation}
where we allow two contributions to the flux of magnetic helicity from the
fluctuating field $\meanFFFFf$: an advective flux due to the wind,
$\meanFFFFf^{\rm w}=\meanhf\meanU_{\rm w}$,
and a turbulent--diffusive flux due to turbulence,
modelled by a Fickian diffusion term down the gradient
of $\meanhf$, i.e., $\meanFFFFf^{\rm diff}=-\kappa_h\nab\meanhf$.
Here, $\meanEMF=\overline{\uu\times\bb}$ is the electromotive force
of the fluctuating field.
The scaling of the terms on the right-hand side with $\Rm$ has been
considered before by \cite{MCCTB10} and \cite{HB10}.
They also drew attention to the fact that, even though $\meanFFFFf$
is gauge-invariant, the time average of $\nab\cdot\meanFFFFf$ is not,
{\em provided} the system is statistically stationary and
$\partial\meanhf/\partial t$ vanishes on average.

In \Fig{presults2b} we show the basic result of \cite{DSGB13}.
As it turns out, below $\Rm=100$, the $2\eta\mu_0\overline{\jj\cdot\bb}$ term
dominates over $\nab\cdot\meanFFFFf$, but because of the different scalings
(slopes being $-1$ and $-1/2$, respectively), the $\nab\cdot\meanFFFFf$ term is
expected to becomes dominant for larger values of $\Rm$ (about 3000).
Unexpectedly, however, $\nab\cdot\meanFFFFf$ becomes approximately
constant already for $\Rm\ga100$ and $2\eta\mu_0\overline{\jj\cdot\bb}$
shows now a shallower scaling (slope $-1/2$).
This means that that the two curves would still cross at a similar value.
Our data suggest, however, that $\nab\cdot\meanFFFFf$ may even rise slightly,
so the crossing point is now closer to $\Rm=1000$.

\section{Solar wind observations}

Solar wind observations provide a good way of determining the
energy spectrum of MHD turbulence.
As already mentioned in \Sec{DynamoGenerated}, recent work by
\cite{BPBP11} provides an explanation of why the kinetic and magnetic
energy spectra have slightly different spectral indices in that the
magnetic energy spectrum is slightly steeper ($\propto k^{-1.6}$) than
that of the kinetic energy ($\propto k^{-1.4}$).
This was found previously by \cite{PRG07}.
Indeed, looking again at \Fig{hyper_512}, it is clear that different
slopes of kinetic and magnetic energy spectra is a consequence of the
super-equipartition just below $k=\kf$ and the subsequent trend toward
equipartition for larger values of $k$.

\begin{figure}[t!]\begin{center}
\includegraphics[width=.8\columnwidth]{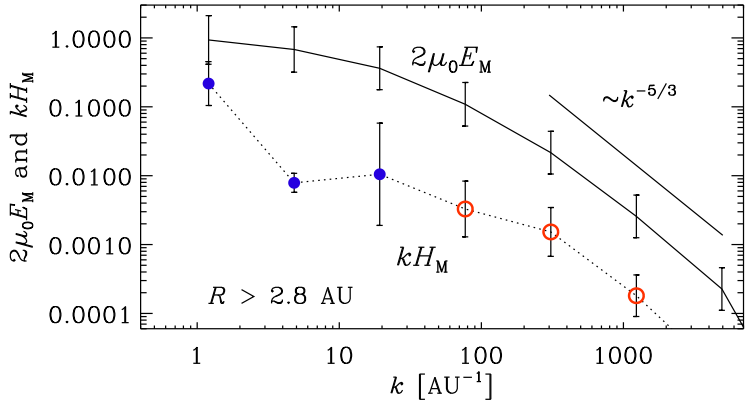}
\end{center}\caption[]{
Magnetic energy and helicity spectra, $2\mu_0 E_{\rm M}(k)$
and $kH_{\rm M}(k)$, respectively, for two separate distance intervals.
Furthermore, both spectra are scaled by $4\pi R^2$ before averaging
within each distance interval above $2.8\AU$.
Filled and open symbols denote negative and positive values of
$H_{\rm M}(k)$, respectively.
Adapted from \cite{BSBG11}.
}\label{rmeanhel4b}\end{figure}

Solar wind observations have long been able to provide estimates
about the magnetic helicity spectrum \citep{MGS82}.
We recall that, even though the magnetic helicity is gauge-dependent,
its spectrum is not.
Technically, this is because the computation of the spectrum
involves an integration over all space.
In practice, this is not possible, of course.
However, by making use of statistical homogeneity and the Taylor
hypothesis of the equivalence of spatial and temporal Fourier spectra,
\cite{MGS82} were able to express the magnetic helicity spectrum as
\begin{equation}
H(k_R)=4\,\mbox{Im}(\hat{B}_T\hat{B}_N^\star)/k_R,
\label{HkR}
\end{equation}
where $\hat{B}_T(k_R)$ and $\hat{B}_N(k_R)$ are the Fourier transforms
of the two magnetic field components perpendicular to the radial
direction away from the Sun, and $(R,T,N)$ refers to the
components of a locally Cartesian heliospheric coordinate system.
Here, $k_R$ is the wavenumber, which is related to the temporal
frequency via $\omega=u_R k_R$, where $u_R\approx800\km\s^{-1}$
is the wind speed at high heliographic latitudes.
Note that in \Eq{HkR}, the expression for $H(k_R)$ is manifestly
gauge-invariant.

Most spacecrafts have probed low heliographic latitudes, where the
helicity is governed by fluctuations around zero.
In recent years, however, it has been possible to estimate the
magnetic helicity spectrum also at high heliographic latitudes using
data from the Ulysses spacecraft that flew in a near-polar orbit.
However, even at high heliographic latitudes the magnetic helicity
is still strongly fluctuating and a clear sign of
magnetic helicity can only be seen by averaging the spectra over broad,
logarithmically spaced wavenumber bins; see \Fig{rmeanhel4b}.
One can define the {\em relative} spectral magnetic helicity,
$2\mu_0\EM(k)/k\HM(k)$, which is a non-dimensional quantity
between $-1$ and $1$.
It turns out that it is just a few percent.
Nevertheless, the magnetic helicity is negative at
large scales (small wavenumbers, $k<30\AU^{-1}$ corresponding to
frequencies below $0.03\mHz$) and positive at smaller scales
(large wavenumbers); see \cite{BSBG11}.
This agrees, at least qualitatively, with earlier results by \cite{SB93}
that at low frequencies the magnetic helicity is negative in the northern
hemisphere.
At much higher frequencies (beyond $100\mHz$), positive magnetic helicity
in the northern hemisphere has now also been found by \cite{PG11}.

When comparing with numerical simulations, it should be noted
that virtually all observed spectra are based on one-dimensional
measurements, while those of numerical simulations are based on
the full three-dimensional velocity field.
The two are related to each other via
\EQ
E_{\rm M}^{\rm1D}(k_R)=\int_{k_R}^\infty E_{\rm M}^{\rm3D}(k)\,\dd\ln k,
\label{EMto1D}
\EN
\EQ
H_{\rm M}^{\rm1D}(k_R)=\int_{k_R}^\infty H_{\rm M}^{\rm3D}(k)\,\dd\ln k.
\label{HMto1D}
\EN
This transformation is well known for the energy spectrum
\citep[cf.,][]{TL72,DHYB03}, and was recently generalized
to the case with helicity \citep{BSBG11}.
The resulting one- and three-dimensional spectra agree in the case of
pure power laws, but near the dissipative cutoff wavenumbers there is
a sharp departure from power law behavior.
This is significant in view of the fact that energy spectra
of three-dimensional simulations indicate the presence of a
so-called bottleneck effect \citep{Fal94}.
This corresponds to an uprise of the compensated energy spectrum,
$k^{5/3}\EK(k)$, near the dissipative cutoff wavenumber
$k_\nu=\bra{\ww^2/\nu^2}^{1/4}$.
This bottleneck effect is much weaker or absent in one-dimensional
spectra \citep{DHYB03,BL09}.
The bottleneck might therefore be a real effect.
Although it happens at such small scales that is should not be
astrophysically significant, it does play a role in three-dimensional
simulations and can lead to effects whose astrophysical significance
needs to be assessed carefully in view of the fact that the growth rate
of small-scale dynamos depends on the shape of the spectrum at the
resistive scale; see \Sec{SmallScaleDynamoLowPm}.

\section{Concluding comments}

The last decades have been marked by important advances in our
understanding of MHD turbulence.
To a substantial degree this happened as numerical simulations became
capable of producing high resolution MHD cubes.
Therefore, MHD turbulence became a theory that can be tested.
As a result of both analytical and numerical studies, as well as
observational measurements of turbulence, the GS95 model of MHD turbulence
has been established as the most promising model.
While we believe that the model is not complete in detail (e.g.\ in
terms of intermittency), it is able to describe the astrophysically
important properties of turbulence, for instance, the scale dependence
of local anisotropy important for cosmic ray propagation, and the magnetic
field wandering important for heat transfer and magnetic reconnection.

The physical ideas of the GS95 model have been extended and applied to
successfully describing compressible MHD turbulence.
It has been shown that the low coupling of fast and Alfv\'en modes allows
the independent treatment of the Alfv\'enic cascade, which is very
important; see \Sec{CompressibleMHDturbulence}.
Indeed, it allows one to use the GS95 scaling for describing Alfv\'enic
modes in moderately compressible fluids, which is of major astrophysical
significance.

However, there are many issues that require further studies.
Those include the properties of highly compressible, highly supersonic
MHD turbulence, scaling and properties of fast modes etc.
Last but not the least, more work is required for the highly debated
subject of imbalanced turbulence.
The corresponding studies call for extensive numerical efforts to test
the existing theories.
We hope that many of these currently controversial issues will be solved
in the near future.
This has important applications for turbulent dynamos of all sorts.
It is now clear that nonlinear turbulent dynamos work also at small
magnetic Prandtl numbers, even though the excitation conditions for
kinematic dynamos become prohibitively high at low magnetic Prandtl
numbers of around 0.1.
As discussed in \Sec{SmallScaleDynamoLowPm}, the reason for this has
meanwhile been identified as the bottleneck effect in turbulence.

Large-scale dynamos are affected by similar subtleties.
They are in particular subject to the possibility of catastrophic quenching,
which means that dynamos and their underlying turbulent transport
coefficients remain dependent on the magnetic Reynolds numbers.
Astrophysical dynamos are believed to be independent of $\Rm$, but we now
know that most dynamos in DNS are probably not yet in that regime,
but there is not much doubt that such a regime exists that is independent
of the magnetic Reynolds number.
In practice, this is accomplished by magnetic helicity fluxes.
Regarding solar and stellar dynamo theory, the reason for equatorward
migration of magnetic activity belts is still not understood.
This is an example where simulations might now lead the way toward
explaining the observed solar behavior, but more progress is needed
to fully understand the physics behind the behavior seen in simulations.

\begin{acknowledgements}
We thank Andre Balogh for providing an inspiring atmosphere at the
International Space Science Institute in Bern in 2012, which has led
to new collaborations and scientific progress.
Computing resources were provided by the Swedish National Allocations Committee
at the Center for Parallel Computers at the Royal Institute of Technology in
Stockholm and the High Performance Computing Center North in Ume{\aa}.
This work was supported in part by the European Research Council
under the AstroDyn Research Project No.\ 227952
and the Swedish Research Council under the project grants 621-2011-5076 and 2012-5797.
AL acknowledges the support of the NSF grant AST-1212096, the NASA grant 
NNX09AH78G, the Vilas Associate Award as well as the support of the 
NSF Center for Magnetic Self-Organization. In addition, AL thanks the International
Institute of Physics (Natal, Brazil) for its hospitality during the work on this review.

\end{acknowledgements}


\end{document}

%% file: macros.tex
\newcommand{\EQ}{\begin{equation}}
\newcommand{\EN}{\end{equation}}
\newcommand{\EQA}{\begin{eqnarray}}
\newcommand{\ENA}{\end{eqnarray}}
\newcommand{\eq}[1]{(\ref{#1})}

\newcommand{\Eq}[1]{Equation~(\ref{#1})}
\newcommand{\Eqs}[2]{Equations~(\ref{#1}) and~(\ref{#2})}

\newcommand{\Sec}[1]{Section~\ref{#1}}

\newcommand{\Fig}[1]{Figure~\ref{#1}}

\newcommand{\Tab}[1]{Table~\ref{#1}}

\newcommand{\bra}[1]{\langle #1\rangle}

{}
{}
{}
\newcommand{\meanFFFFf}{\overline{\mbox{\boldmath ${\cal F}$}}_{\rm f}{}}{}
\newcommand{\meanemf}{\overline{\cal E} {}}

{}
{}
\newcommand{\meanEMF}{\overline{\mbox{\boldmath ${\cal E}$}}{}}{}
{}
{}
{}
{}
{}
{}
\newcommand{\meanBB}{\overline{\mbox{\boldmath $B$}}{}}{}
{}
{}
\newcommand{\meanFFf}{\overline{\mbox{\boldmath $F$}}_{\rm f}{}}{}
{}
{}
{}
{}
{}
\newcommand{\meanJJ}{\overline{\mbox{\boldmath $J$}}{}}{}
\newcommand{\meanKK}{\overline{\mbox{\boldmath $K$}}{}}{}
\newcommand{\meanUU}{\overline{\mbox{\boldmath $U$}}{}}{}

{}
{}
{}

\newcommand{\meanB}{\overline{B}}

\newcommand{\meanC}{\overline{C}}

\newcommand{\meanhf}{\overline{h}_{\rm f}}

\newcommand{\meanU}{\overline{U}}

{}

{}
{}

\newcommand{\eee}{\hat{\mbox{\boldmath $e$}} {}}

\newcommand{\OOO}{\hat{\mbox{\boldmath $\Omega$}} {}}

\newcommand{\xx}{\mbox{\boldmath $x$} {}}

\newcommand{\ww}{\mbox{\boldmath $w$} {}}

\newcommand{\kk}{\bm{k}}

\newcommand{\uu}{\mbox{\boldmath $u$} {}}
\newcommand{\UU}{\mbox{\boldmath $U$} {}}

\newcommand{\bb}{\mbox{\boldmath $b$} {}}
\newcommand{\BB}{\mbox{\boldmath $B$} {}}

\newcommand{\jj}{\mbox{\boldmath $j$} {}}
\newcommand{\JJ}{\mbox{\boldmath $J$} {}}

\newcommand{\AAA}{\mbox{\boldmath $A$} {}}
\newcommand{\aaaa}{\mbox{\boldmath $a$} {}}

\newcommand{\ff}{\mbox{\boldmath $f$} {}}

\newcommand{\WW}{\mbox{\boldmath $W$} {}}

\newcommand{\grav}{\mbox{\boldmath $g$} {}}
\newcommand{\nab}{\mbox{\boldmath $\nabla$} {}}
\newcommand{\OO}{\bm{\Omega}}

\newcommand{\ggamma}{\mbox{\boldmath $\gamma$} {}}

\newcommand{\RRRR}{\mbox{\boldmath ${\sf R}$} {}}

\newcommand{\ii}{{\rm i}}

\newcommand{\dd}{{\rm d} {}}

\newcommand{\const}{{\rm const}  {}}

\def\degr{\hbox{$^\circ$}}
\def\la{\mathrel{\mathchoice {\vcenter{\offinterlineskip\halign{\hfil
$\displaystyle##$\hfil\cr<\cr\sim\cr}}}
{\vcenter{\offinterlineskip\halign{\hfil$\textstyle##$\hfil\cr<\cr\sim\cr}}}
{\vcenter{\offinterlineskip\halign{\hfil$\scriptstyle##$\hfil\cr<\cr\sim\cr}}}
{\vcenter{\offinterlineskip\halign{\hfil$\scriptscriptstyle##$\hfil\cr<\cr\sim\cr}}}}}
\def\ga{\mathrel{\mathchoice {\vcenter{\offinterlineskip\halign{\hfil
$\displaystyle##$\hfil\cr>\cr\sim\cr}}}
{\vcenter{\offinterlineskip\halign{\hfil$\textstyle##$\hfil\cr>\cr\sim\cr}}}
{\vcenter{\offinterlineskip\halign{\hfil$\scriptstyle##$\hfil\cr>\cr\sim\cr}}}
{\vcenter{\offinterlineskip\halign{\hfil$\scriptscriptstyle##$\hfil\cr>\cr\sim\cr}}}}}
\def\Co{\mbox{\rm Co}}

\def\Pm{\mbox{\rm Pr}_M}
\def\Rm{\mbox{\rm Re}_M}

\def\Rmc{\mbox{\rm Re}_{M,{\rm crit}}}
\def\Rey{\mbox{\rm Re}}
\def\Pe{\mbox{\rm Pe}}
\def\Co{\mbox{\rm Co}}

\def\EK{E_{\rm K}}
\def\EM{E_{\rm M}}
\def\EC{E_{\rm C}}

\def\cs{c_{\rm s}}

\def\qp{q_{\rm p}}

\def\kf{k_{\rm f}}
\def\kfz{k_{\rm f0}}
\def\HM{H_{\rm M}}
\def\HK{H_{\rm K}}
\def\epsf{\epsilon_{\rm f}}

\def\kM{k_{\rm M}}

\def\Brms{B_{\rm rms}}

\def\urms{u_{\rm rms}}

\def\kappat{\kappa_{\rm t}}
\def\kappaT{\kappa_{\rm T}}
\def\kappatz{\kappa_{\rm t0}}

\def\etat{\eta_{\rm t}}
\def\etatz{\eta_{\rm t0}}

\def\etaT{\eta_{\rm T}}

\def\Beq{B_{\rm eq}}

\def\half{{\textstyle{1\over2}}}

\newcommand{\mHz}{\,{\rm mHz}}

\newcommand{\s}{\,{\rm s}}

\newcommand{\km}{\,{\rm km}}

\newcommand{\AU}{\,{\rm AU}}

\newcommand{\yan}[5]{, #5, {Astron.\ Nachr.\ }{\bf #2}, #3-#4 (#1)}

\newcommand{\yana}[5]{, #5, {Astron.\ Astrophys.\ }{\bf #2}, #3-#4 (#1)}
\newcommand{\yanaN}[4]{, #4, {Astron.\ Astrophys.\ }{\bf #2}, #3 (#1)}

\newcommand{\yass}[5]{, #5, {Astrophys.\ Spa.\ Sci.\ }{\bf #2}, #3-#4 (#1)}
\newcommand{\yssr}[5]{, #5, {Spa.\ Sci.\ Rev.\ }{\bf #2}, #3-#4 (#1)}

\newcommand{\ysph}[5]{, #5, {Solar Phys.\ }{\bf #2}, #3-#4 (#1)}
\newcommand{\dsph}[3]{, #2, {Solar Phys.}, DOI:#3 (#1)}
\newcommand{\yjetp}[5]{, #5, {Sov.\ Phys.\ JETP }{\bf #2}, #3-#4 (#1)}

\newcommand{\ysov}[5]{, #5, {Sov.\ Astron.\ }{\bf #2}, #3-#4 (#1)}
\newcommand{\ysovl}[5]{, #5, {Sov.\ Astron.\ Lett.\ }{\bf #2}, #3-#4 (#1)}
\newcommand{\ymn}[5]{, #5, {Monthly Notices Roy.\ Astron.\ Soc.\ }{\bf #2}, #3-#4 (#1)}
\newcommand{\ymnS}[5]{, #5 {Monthly Notices Roy.\ Astron.\ Soc.\ }{\bf #2}, #3-#4 (#1)}

\newcommand{\yjfm}[5]{, #5, {J.\ Fluid Mech.\ }{\bf #2}, #3-#4 (#1)}
\newcommand{\ypr}[5]{, #5, {Phys.\ Rev.\ }{\bf #2}, #3-#4 (#1)}
\newcommand{\ypre}[5]{, #5, {Phys.\ Rev.\ E }{\bf #2}, #3-#4 (#1)}
\newcommand{\yprdN}[4]{, #4, {Phys.\ Rev.\ D }{\bf #2}, #3 (#1)}
\newcommand{\ypreN}[4]{, #4, {Phys.\ Rev.\ E }{\bf #2}, #3 (#1)}
\newcommand{\ypreNS}[4]{, #4 {Phys.\ Rev.\ E }{\bf #2}, #3 (#1)}
\newcommand{\yprl}[5]{, #5, {Phys.\ Rev.\ Lett.\ }{\bf #2}, #3-#4 (#1)}
\newcommand{\yprlN}[4]{, #4, {Phys.\ Rev.\ Lett.\ }{\bf #2}, #3 (#1)}

\newcommand{\yprs}[5]{, #5, {Proc.\ Roy.\ Soc.\ Lond.\ }{\bf #2}, #3-#4 (#1)}

\newcommand{\yapj}[5]{, #5, {Astrophys.\ J.\ }{\bf #2}, #3-#4 (#1)}
\newcommand{\sapj}[3]{, #2, {Astrophys.\ J.}, submitted, arXiv:#3 (#1)}
\newcommand{\sapjl}[3]{, #2, {Astrophys.\ J.\ Lett.}, submitted, arXiv:#3 (#1)}

\newcommand{\yapjN}[4]{, #4, {Astrophys.\ J.\ }{\bf #2}, #3 (#1)}
\newcommand{\yapjlN}[4]{, #4, {Astrophys.\ J.\ Lett.\ }{\bf #2}, #3 (#1)}
\newcommand{\yapjS}[5]{, #5, {Astrophys.\ J.\ }{\bf #2}, #3-#4 (#1)}

\newcommand{\yapjs}[5]{, #5, {Astrophys.\ J.\ Suppl.\ }{\bf #2}, #3-#4 (#1)}
\newcommand{\yapjl}[5]{, #5, {Astrophys.\ J.\ Lett.\ }{\bf #2}, #3-#4 (#1)}
\newcommand{\yjpp}[5]{, #5, {J.\ Plasm.\ Phys.\ }{\bf #2}, #3-#4 (#1)}

\newcommand{\yaraa}[5]{, #5, {Ann.\ Rev.\ Astron.\ Astrophys.\ }{\bf #2}, #3-#4 (#1)}

\newcommand{\yanf}[5]{, #5, {Ann.\ Rev.\ Fluid Dyn.\ }{\bf #2}, #3-#4 (#1)}
\newcommand{\ypf}[5]{, #5, {Phys.\ Fluids }{\bf #2}, #3-#4 (#1)}
\newcommand{\ypfN}[4]{, #4, {Phys.\ Fluids }{\bf #2}, #3 (#1)}

\newcommand{\ygafd}[5]{, #5, {Geophys.\ Astrophys.\ Fluid Dynam. }{\bf #2}, #3-#4 (#1)}

\newcommand{\yjour}[6]{, #6, {#2} {\bf #3}, #4-#5 (#1)}
\newcommand{\yjourN}[5]{, #5, {#2} {\bf #3}, #4 (#1)}

\newcommand{\djour}[4]{, #3, {#2}, DOI: #4 (#1)}

\newcommand{\yproc}[7]{, #4, In {#5} (ed.\ #6), pp.\ #2-#3.\ #7 (#1)}

\newcommand{\ybook}[3]{, {#2}.\ #3 (#1)}